\documentclass[review,onefignum,onetabnum]{siamonline220329}

\usepackage{lipsum}
\usepackage{amsfonts}
\usepackage{graphicx}
\usepackage{epstopdf}
\usepackage{algorithmic}
\ifpdf
  \DeclareGraphicsExtensions{.eps,.pdf,.png,.jpg}
\else
  \DeclareGraphicsExtensions{.eps}
\fi
\usepackage{bm}
\usepackage{subcaption}

\usepackage{enumitem}
\setlist[enumerate]{leftmargin=.5in}
\setlist[itemize]{leftmargin=.5in}

\newsiamremark{remark}{Remark}
\newsiamremark{hypothesis}{Hypothesis}
\crefname{hypothesis}{Hypothesis}{Hypotheses}
\newsiamthm{claim}{Claim}

\headers{Assessment of CRAN for learning wave propagation}{W. Mallik, R. K. Jaiman, and J. Jelovica}

\title{Assessment of convolutional recurrent autoencoder network for learning wave propagation
}

\author{Wrik Mallik\thanks{James Watt School of Engineering,  University of Glasgow, Glasgow, G12 8QQ, UK 
  (\email{wrik.mallik@glasgow.ac.uk}).}
\and Rajeev K. Jaiman\thanks{Department of Mechanical Engineering,  The University of British Columbia, Vancouver, BC V6T 1Z4, Canada 
  (\email{rjaiman@mech.ubc.ca}).}
\and Jasmin Jelovica\thanks{Departments of Mechanical and Civil Engineering,  The University of British Columbia, Vancouver, BC V6T 1Z4, Canada 
  (\email{jasmin.jelovica@ubc.ca}).}}

\usepackage{amsopn}

\ifpdf
\hypersetup{
  pdftitle={Assessment of CRAN for learning wave propagation},
  pdfauthor={Wrik Mallik, Rajeev K. Jaiman, and Jasmin Jelovica}
}
\fi

\usepackage{mathtools}
\usepackage{xcolor}
\begin{document}
\nolinenumbers

\maketitle

\begin{abstract}
It is challenging to construct generalized physical models of wave propagation in nature owing to their complex physics as well as widely varying environmental parameters and dynamical scales. In this article, we present the convolutional autoencoder recurrent network (CRAN) as a data-driven model for learning wave propagation phenomena. The CRAN consists of a convolutional autoencoder for learning low-dimensional system representation and a long short-term memory recurrent neural network for the system evolution in low dimension. 
We show that the convolutional autoencoder significantly outperforms the dimension-reduction of complex wave propagation phenomena via projection-based methods as it can directly learn subspaces resembling wave characteristics. On the other hand, the projection-based modes are restricted to the Fourier subspace. Geometric priors of the convolutional autoencoder enabling selective scale separation of complex wave dynamics further enhance its dimension-reduction capability. We also demonstrate that geometric priors such as translation equivariance and translational invariance of the convolutional autoencoder enable generalized learning of low-dimensional maps. Thus, the composite CRAN model connecting the convolutional autoencoder with a long short-term memory network specially designed for autoregressive modeling can perform generalized wave propagation prediction over the desired time horizon. Numerical experiments display 90\% mean structural similarity index measure of CRAN predictions compared to true solutions for out-of-training cases, and less than 10\% pointwise $L_1$ error for most cases, verifying such generalization claims. Finally, the CRAN predictions offer similar wave characteristic patterns to the target solutions indicating not only their generalization but also their kinematical consistency.
\end{abstract}

% REQUIRED

% REQUIRED

%\section{Overview: presently for internal use}
%\begin{itemize}
    %\item This article connects the inductive bias (translation equivariance and invaraince, scale separation, etc.) of CNNs presented by Bronstein, Manosur (lazy learning) to data-driven learning of wave propagation by interpreting the dimension reduction and generalization of convolutional autoencoders. Previous well-known applications discuss dimension reduction (Carlberg et al., Manzoni et al, etc.) but ignores the generalization and inductive bias of geometric DL models. 
    %\item We show that the flexibility of CNN basis functions (directly learning wave charactersitics) and their selective scale separation (via pooling operation) enables them to reduce high-dimensional wave propagation phenomena with discontinuities to a low-dimensional set of latent states. POD on the other hand employ Fourier modes for learning a low-dimensional representation leading to a slow decay of Kolmogorov $n$-width.  
    
    %\item Wave propagation with spatially varying sources can be considered as an equivariant map from the wave sources to the resulting wave charactersitics.  Thus, the ivariance and equivairance to translation endowed to CNN autoencoders thus help in generalization. Generalization is demonstrated by correctly predicting various scales of wave characteristic features for source locations not included in the training set.
    
    %\item Waves are propagated along characteristic lines. Generalization of wave characteristics demonstrates kinematical consistency.
%\end{itemize}

\section{\label{sec:intro} Introduction}
% Motivation of our research
Complex wave propagation phenomena can be observed in various forms around us ranging from seismic activities to ocean acoustics. It is challenging to develop generalized physical models for predicting large-scale wave propagation in nature, owing to a complex multiphysics phenomenon with widely varying physical scales. Environmental wave propagation phenomena are analyzed via techniques of varying fidelity according to the nature of the physical problem investigated. As a result, high-fidelity (e.g., finite difference, finite element) and low-fidelity (e.g., ray tracing, wavenumber integration) numerical techniques are routinely used according to specific situations and requirements \cite{jensen2011computational}. However, it would be highly beneficial to construct a generalized model that could unify various analysis techniques over a wide range of parameters together with measurement data. 

One can attempt to develop generalized models of wave propagation via data-driven techniques. Being data-driven, they rely only on the quality of the observed data and remain agnostic to the how the data is obtained. Therefore, they could potentially encompass data with varying resolutions that are obtained across a wide range of parameters. Such data-driven models should ideally enable an offline-online application strategy. During the offline stage, the data-driven model can be trained to learn a low-dimensional representation of the system from the high-dimensional physical data. During the online phase, the reduced-dimensional data-driven model can provide real-time and scalable predictions for various applications including decision making, control and optimization.

% background on data-driven projection-based ROMs
Various data-driven techniques have been historically employed for learning dynamical systems exhibiting nonlinear and multi-scale behavior. Projection-based methods such as proper orthogonal decomposition (POD) are one such popularly used technique, which employ linear trial subspaces to learn the low-dimensional representation of the system. The worst-case error from such optimal linear subspaces can be represented by the Kolmogorov $n$-width, which is defined as
\begin{equation*}
    d_n\left(\mathcal{M}\right) \coloneqq \inf_{V_n} \sup_{u \in \mathcal{M}} \inf_{v_n \in V_n} {\lVert u - v_n \rVert}.
\end{equation*}
Here $\mathcal{M}$ is the solution manifold over all time and spatial parameters, and $V_n$ is the set of all possible linear trial subspaces of dimension $n$. The employment of a linear trial subspace can be justified for certain physical problems like diffusion-dominated problems, which exhibit a fast decaying Kolmogorov $n$-width \cite{bachmayr2017kolmogorov}. However, many advection-dominated problems, including wave propagation, exhibit a slowly decaying Kolmogorov $n$-width \cite{greif2019decay,taddei2020registration,lee2020model}, indicating that projection-based methods will not be efficient dimension reduction techniques. This has motivated the investigation of nonlinear dimension reduction approaches, which can directly learn nonlinear manifolds from snapshots of high-dimensional data. Some such approaches not only provide the low-dimensional representation of the high-dimensional data but also provide the map to retrieve the high-dimensional data back from the low-dimensional representation. These include self-organizing maps \cite{kohonen1982self}, kernel principal component analysis \cite{scholkopf1998nonlinear}, diffeomorphic dimensionality reduction approaches \cite{walder2008diffeomorphic, taddei2020registration} and deep learning-based model, the autoencoders \cite{hinton2006reducing}. In this study, we restrict ourselves to autoencoders.

Autoencoders are neural network architectures comprising an encoder and a decoder network. The encoder can compress high-dimensional physical data to a much lower-dimensional latent states via a composition of nonlinear neural network activation functions. The decoder subsequently generates a nonlinear compositional map to retrieve the high-dimensional physical data from the low-dimensional latent states. Furthermore, unlike projection-based models, the basis functions of the encoder and decoder are selected from an affine subspace via a heuristic process. Autoencoders have been recently demonstrated as a completely non-intrusive data-driven dimension reduction approach for various physical phenomena. These include low-dimensional modeling of wall turbulence \cite{milano2002neural}, dimension reduction of advection-dominated dynamical systems \cite{gonzalez2018deep, lee2020model, xu2020multi, fotiadis2020comparing, maulik2021reduced, mallik2022predicting} or even learning approximate invariant subspaces of the Koopman operator \cite{takeishi2017learning, lusch2018deep, otto2019linearly}. 

There has been a recent shift in the deep-learning-based dimension reduction studies from vanilla autoencoders comprised of fully connected neural networks to convolutional neural network-based autoencoders. Several such application of convolutional autoencoders (CAEs) for learning advection-dominated problems in low-dimensions exist \cite{gonzalez2018deep, lee2020model, gonzalez2018deep,xu2020multi,fotiadis2020comparing,bukka2021assessment}, and their superior dimension reduction capacity compared to fully projection-based models have also been reported for advection-dominated problems \cite{gonzalez2018deep,lee2020model,fresca2021comprehensive}. However, the reason for the such choice of the deep-learning model has not been fully discussed apart from the fact that CAEs require fewer training parameters \cite{lee2020model}. Also, these articles do not physically interpret how CAEs learn the low-dimensional representation of the high-dimensional physical system, although similar physical interpretability of dimension reduction via projection-based approaches \cite{otto2019linearly} can be readily obtained from the highest energy projected modes. 

It is also important to note that data-driven scalability relies not only on their dimensional reduction capacity but also on generalization over parameter space. Thus, such generalized learning of dynamical systems via data-driven models is desired over a wide range of parameters. However, such generalization of advection-dominated dynamical systems via CAEs is demonstrated in only a few articles \cite{fotiadis2020comparing,xu2020multi,maulik2021reduced,mallik2022predicting}. Moreover, these studies do not discuss why CAEs are able to demonstrate such generalization. Similarly, no physical interpretability is provided to explain how such generalized learning of the dynamical system is achieved.

In this article, we physically interpret both the low-dimensional representation of dynamical systems learned by CAEs and also the generalized learning of dynamical systems. We connect the inductive bias of convolutional neural networks \cite{bronstein2017geometric, bronstein2021geometric} obtained via various geometric priors to their dimension-reduction and generalized learning capabilities. These geometric priors like translational equivariance and translational invariance and stability to deformations separate the convolutional neural nets from nonlinear regression models like fully connected neural networks. To demonstrate such interpretability, we specifically select wave propagation phenomena, showing a slow decay of Kolmogorov $n$-width via projection-based dimension reduction methods. We investigate the feature maps of various encoder layers to interpret both why and how the CAEs can both efficiently reduce high-dimensional physical data of convection-dominated problems on a low-dimensional manifold and also learn generalized wave propagation with spatially varying parameters. 

To complete the data-driven dynamical system prediction in recent studies, CAEs have been integrated with either numerical time-integration schemes \cite{lee2020model, fresca2021comprehensive}, or various neural network-based temporal propagation schemes to learn the dynamics \cite{xu2020multi,gonzalez2018deep,bukka2021assessment,fotiadis2020comparing}. Here we adopt the second path and integrate the CAE to long-short term memory (LSTM) recurrent neural networks to temporally evolve the low-dimensional latent states learned by the encoder. The composite encoder-propagate-decoder network is called the convolutional recurrent autoencoder network (CRAN). The CRAN network is employed to demonstrate generalized learning of wave propagation. The CRAN has been employed in the past for learning complex flow dynamics around bluff bodies \cite{bukka2021assessment}. However, here we enhance the propagator of the previous CRAN model by including LSTM architectures with autoregressive dynamics modeling to improve its prediction of wave propagation dynamics. Subsequently, the CRAN is employed for generalized learning of wave propagation with a spatially varying wave initial conditions. The application of CRAN for generalized data-driven learning of wave propagation and the physical interpretation of the nonlinear dimension reduction and generalized wave propagation learning mechanism is presented here for the first time.

% motivation for using Ml/DL techniques
%In recent years, deep learning (DL) architectures based on neural networks have been increasingly used as data-driven models for mechanistic problems \cite{hsieh2009machine, bergen2019machine, reichstein2019deep}. Recent numerical experiments \cite{mansour2019deep} indicate that DL models tend to prioritize the learning of inherent simpler features of the data over the complex data patterns. Such behavior is observed even when the training data is noisy. Such learning of simple laws governing the data aligns perfectly with data of mechanistic origin, which are governed by a few fundamental physical principles (e.g., Newton's laws of motion). This underscores the motivation for approximating the mechanistic systems with data-driven DL models.

\section{\label{sec:2} Data-driven learning of wave propagation}
Here we begin with formulating a general data-driven learning of dynamical systems, which will eventually be applied for learning wave propagation. Any general parametric partial differential equation depending on spatial coordinates $\bm{\mathrm{x}} \in \mathbb{R}^i \left(i=1,2,3\right)$ and on the time $t \in I \subset \mathbb{R}^+$ for real-valued parameter $\Phi \in \mathbb{R}^P$ can be written as
\begin{equation}
    \label{eq:infdimpde}
    \frac{\partial}{\partial t}\bm{\mathrm{U}} \left(\bm{\mathrm{x}}, t;\Phi \right) = \mathcal{F}\left(\bm{\mathrm{U}} \left(\bm{\mathrm{x}}, t;\Phi \right); \Phi \right),
\end{equation}
subject to any general initial and boundary conditions. $\mathcal{F}$ is any general nonlinear operator governing the dynamics of the observable $\bm{\mathrm{U}}$. In most practical situations, a closed-form solution to Eq. (\ref{eq:infdimpde}) does not exist and we obtain a discretized representation of our observable (e.g., numerical solutions or experimental measurements) in both space and time. Under such situations we obtain a discrete representation of Eq. (\ref{eq:infdimpde}) as 
\begin{equation}
    \label{eq:discpde}
    \bm{\mathrm{U}}_{M,k+1} \left(\bm{\mathrm{x}}_M, t_{k+1};\Phi \right) = \mathcal{F}_{M,K}\left(\bm{\mathrm{U}}_{M,k} \left(\bm{\mathrm{x}}_M, t_k;\Phi \right);\Phi \right), \quad k=0,1,\ldots, K-1,
\end{equation}
where $\bm{\mathrm{U}}_{M,k+1} \in \mathbb{R}^M$. $M$ represents the spatial discretization and $K$ represent the discrete time steps over a given time interval. The discrete operator $\mathcal{F}_{M,K}$ depends on the exact form of Eq. (\ref{eq:infdimpde}) and thus provides a forward map from the causality (initial and boundary conditions, system properties and dynamics) to the effects (final solution). 

The objective of the data-driven learning is to obtain an inverse map $\mathcal{G}$ such that we can obtain the evolution of our observable from 
\begin{equation}
    \label{eq:datadrivenpde}
    \bm{\mathrm{U}}_{M,k+1} \left(\bm{\mathrm{x}}_M, t_{k+1};\Phi \right) = \mathcal{G}\left(\bm{\mathrm{U}}_{M,k} \left(\bm{\mathrm{x}}_M, t_k;\Phi \right); \bm{\mathrm{\Theta}} \right), \quad k=0,1,\ldots, K-1.
\end{equation}
$\mathcal{G}$ is dependent on the parameters $\bm{\mathrm{\Theta}}$, which are learned offline from a set of known observables via a data-driven learning technique. Subsequently, $\mathcal{G}$ is autoregressively applied in Eq. \ref{eq:datadrivenpde} over $K-1$ time-steps into the horizon once it is initiated with $\bm{\mathrm{U}}_{M,0}$. Thus, the causality ($\mathcal{G}$) is learned from the effects (known observables). We desire the inverse map to provide a unified model of the system across various sources of data and system parameters. Also, the dynamical system operates on high-dimensional spatial data (i.e., $M$ is large) and a large parameter set $\Phi$. We want to learn the data-driven model on a manifold whose dimensionality is much lower than the physical space on which the dynamical system operates. According to the manifold hypothesis, such low-dimensional manifolds remain embedded within the high-dimensional space representing the physical data. Therefore, identifying such low-dimensional manifolds via data-driven models is possible with the correct choice of the data-driven model.

In this research, we will specifically investigate data-driven learning of wave propagation governing underwater noise radiations in the ocean. Such noise radiations in the ocean environment can be considered as a wave propagation phenomena with the assumptions that the fluid is isotropic and homogeneous, viscous stresses are negligible, the process is adiabatic, and the spatial variations of the ambient pressure, density and temperature are relatively small. Under these assumptions, physical quantities like density and pressure can be expressed as a sum of their steady values and unsteady fluctuations of much smaller amplitude. Using the aforementioned assumptions, the conservation of mass and momentum, and the equation of state of the fluid, the propagation of pressure fluctuations for $\bm{\mathrm{x}} = \left( x, y, z \right)$ can be represented as a second-order hyperbolic partial differential equation,
\begin{equation}
    \label{eq:2ndorder_wave_PDE}
     \nabla^2 p - \frac{1}{{c_o}^2} \frac{\mathrm{\partial^2} p}{{\mathrm{\partial} t}^2} = q(\bm{\mathrm{x}},t;\bm{\mathrm{x_0}},t_0), \quad \left(t,\bm{\mathrm{x}}\right) \in I\times \Omega,
\end{equation}
subject to the initial and boundary conditions. Here $\nabla^2 = \frac{\partial^2}{\partial  x^2} + \frac{\partial^2}{\partial  y^2} + \frac{\partial^2}{\partial  z^2}$. $q$ is a source term subject to the source location $\bm{\mathrm{x_0}}$ and initiation point $t_0$. $c_0$ is the constant speed of sound in the medium domain represented by $\Omega$. Thus, equation (\ref{eq:2ndorder_wave_PDE}) can also be expressed in the general form using Eq. (\ref{eq:infdimpde}) where $\bm{\mathrm{U}} = \left(p,\frac{\partial p}{\partial t}\right)$ and $\phi$ represents the set of parameters including boundary conditions, domain and source properties. 

Here, we employ a data-driven approach to learn the propagation of acoustic pressure waves. Following Eq. \ref{eq:datadrivenpde}, the data-driven learning problem can be represented without any loss of generality as
\begin{equation}
    \label{eq:datadrivenacouspres}
    \bm{\mathrm{p}}_{M,k+1} \left(\bm{\mathrm{x}}_M, t_{k+1};\Phi \right) = \mathcal{G}\left(\bm{\mathrm{p}}_{M,k} \left(\bm{\mathrm{x}}_M, t_k;\Phi \right); \bm{\mathrm{\Theta}} \right), \quad k=0,1,\ldots, K-1.
\end{equation}
Here $\bm{\mathrm{p}}$ denotes the acoustic pressure observable, $\phi$ represents the set of parameters (e.g., initial and boundary conditions, domain and source properties) which affect the solution of Eq. (\ref{eq:2ndorder_wave_PDE}), $M$ is spatial discretization and $K$ represent the discrete time steps over a given time interval. As explained previously, the inverse operator $\mathcal{G}$ is learned via data-driven technique, and subsequently employed in an autoregressive manner to obtain the long-time evolution of acoustic pressure in the domain.

\section{\label{sec:3} Data-driven learning methodology}
For any general high-dimensional observed variable of interest $\bm{\mathrm{U}}_{M,k} \in \mathbb{R}^M$, obtained at propagation step $k$, the data-driven learning problem can be formulated as the learning of three operators $\mathcal{E}$, $\mathcal{P}$ and $\mathcal{D}$
\begin{equation}
    \label{eq:data-driven_learning}
    \begin{split}
        \bm{\mathrm{A}}_{n,k}\left(t_k;\Phi \right) =& \mathcal{E}\left(\bm{\mathrm{U}}_{M,k}\left(\bm{\mathrm{x}}_M, t_k;\Phi \right);\bm{\theta}_\mathcal{E} \right),\\
        \bm{\mathrm{A}}_{n,k+1}\left(t_{k+1};\Phi \right) =& \mathcal{P}\left(\bm{\mathrm{A}}_{n,k}\left(t_k;\Phi \right);\bm{\theta}_\mathcal{P}\right),\\
        \tilde{\bm{\mathrm{U}}}_{M,k+1}\left(\bm{\mathrm{x}}_M, t_{k+1};\Phi \right) =&\mathcal{D}\left(\bm{\mathrm{A}}_{n,k+1}\left(t_{k+1};\Phi \right);\bm{\theta}_\mathcal{D}\right).
    \end{split}
\end{equation}
Here $\tilde{\bm{\mathrm{U}}}_{M,k+1} \in \mathbb{R}^M$ and $\bm{\mathrm{A}}_{n,k} \in \mathbb{R}^n$, for all $k=0,1,\ldots,K-1$. The data-driven operator $\mathcal{G}$ can thus be considered a composition of these three operators, 
\begin{equation}
    \label{eq:composite_inverse_map}
    \mathcal{G} = \left(\mathcal{E}\circ\mathcal{P}\circ\mathcal{D};\bm{\mathrm{\Theta}}\right)
\end{equation}
where $\bm{\mathrm{\Theta}} = \left(\bm{\theta}_\mathcal{E},\bm{\theta}_\mathcal{P},\bm{\theta}_\mathcal{D}\right)$. Our objective is to learn $\mathcal{G}$ such that it can operate on a low-dimensional manifold of dimensionality $n$, for a range of parameters $\phi$. In other words, we want $\tilde{\bm{\mathrm{U}}}_{M,k+1}\approx\bm{\mathrm{U}}_{M,k+1}$ while $n\ll M$. Here, the operator $\mathcal{E}$ reduces the high-dimensional dynamical system on a low-dimensional manifold, $\mathcal{P}$ evolves the system along the low-dimensional manifold, and $\mathcal{D}$ expands the evolved low-dimensional system to the original high dimension. We will consider learning the low-dimensional system representation and the evolution of the low-dimensional system as separate learning problems. These are subsequently discussed in this section in detail.

\subsection{Learning low-dimensional representation}
In this study, we will primarily consider the deep learning model, the CAE, to reduce the high-dimensional physical states of the system on a lower-dimension manifold. However, we also compare the dimension reduction of the projection-based POD to the CAE.

\subsubsection{Convolutional autoencoders}
DL-based models, composed of deep neural networks, attempt to find the best low-dimensional nonlinear manifold, which can represent the high-dimensional system. Such nonlinear mapping between low-dimensional and high-dimensional representations becomes especially useful for learning transport problems, where the Kolmogorov $n$-width decays slowly. Here we employ DL models as an autoencoder for obtaining the low-dimensional spatial representation. The autoencoder consists of an encoder, which compresses high-dimensional spatial data to a set of latent states lying on a much lower-dimensional manifold. The decoder can subsequently expand the latent states to their high-dimensional representation. The autoencoder is a combination of the encoder and the decoder and is learned in a semi-supervised manner, as we specify the input and output to the autoencoder, but do not supervise how the latent states are learned.

The specific DL architecture considered for both the encoder and the decoder is Convolutional Neural Network. Convolutional Neural Network is specifically selected as they offer certain geometric priors to the nonlinear encoding map from high-dimensional physical states to the latent states lying on the low-dimensional manifold \cite{bronstein2017geometric}. Similar geometric priors are also endowed to the convolutional decoding map, which expands the low-dimensional latent states to high-dimensional physical states. Such geometric priors enable convolutional neural networks to learn generalized behavior and avoid memorization \cite{mansour2019deep}. Here we briefly discuss the method of learning low-dimensional representation from high-dimensional physical data via convolutional autoencoders. We also discuss how the various geometric priors can help these networks generalize advection-dominated dynamical systems. 

Let us consider a compact $M$-dimensional Euclidean domain $\Omega = {[0,1]}^M \subset \mathbb{R}^M$ on which square-integrable functions $U \in L^2 \left(\Omega \right)$ are defined. We consider a generic semi-supervised learning environment for obtaining an unknown function $a: L^2 \left(\Omega\right) \rightarrow \mathcal{Y}$ that is not observed on a training set, and square-integrable function $\tilde{U}: \mathcal{Y} \rightarrow L^2 \left(\Omega \right)$ that is observed on the training set. Thus, 
\begin{equation}
    {\{U_i,\tilde{U}_i \in L^2 \left(\Omega\right),a_i=a\left(U_i\right) ,\tilde{U}_i=\tilde{U}\left(a_i\right)\}}_{i \in \mathcal{I}},
\end{equation}
where $\mathcal{Y} \in \mathbb{R}^n$. 

The convolutional encoder consists of several convolutional layers of the form $C(\bm{\mathrm{U}})$ which sequentially performs a set of convolution operations of the form $\bm{g}=\kappa_\Lambda(\bm{\mathrm{U}})$ and a point-wise non-linearity $\xi$, acting on a $p$-dimensional input $\bm{\mathrm{U}}(x) = \left( U_1(x), \ldots , U_p(x) \right)$,
\begin{equation}
    C(\bm{\mathrm{U}}) = \xi\left(\kappa_\Lambda(\bm{\mathrm{U}})\right).
\end{equation}
The convolution operation $\kappa_\Lambda(\bm{U})$ operates by applying a set of kernels (or filters) $\Lambda = \left(\lambda_{l,l^{'}}\right), l=1, \ldots, ,q, l^{'} = 1, \ldots, p$,
\begin{equation}
    g_l(x) =  \sum_{l^{'}=1}^p \left(U_{l^{'}}*\lambda_{l,l^{'}}\right)(x),
\end{equation}
producing a $q$-dimensional output $\bm{g}(x) = \left(g_1(x), \ldots, g_q(x)\right)$, often referred to as the feature maps. Here,
\begin{equation}
    \left(U*\lambda\right)(x) = \int_{\Omega} U\left(x-x{'}\right)\lambda\left(x{'}\right)dx{'},
\end{equation}
denotes a standard convolution operation. Similarly, a pooling layer $P(\bm{\mathrm{U}})$ comprising of a downsampling (or pooling) operation $\bm{g}=\gamma_\Lambda(\bm{\mathrm{U}})$, and a point-wise non-linearity $\xi$ can be expressed as
\begin{equation}
    P(\bm{\mathrm{U}}) = \xi\left(\gamma_\Lambda(\bm{\mathrm{U}})\right).
\end{equation}
The pooling operation can be defined as \begin{equation}
    g_l(x) =  \gamma \left(\{U\left(x{'}\right)\}:x{'} \in \mathcal{N}(x) \right), \quad l=1, \ldots, ,q,
\end{equation}
where $\mathcal{N}(x) \subset \Omega$ is a neighbourhood around $x$ and $\gamma$ is a permutation-invariant function such as an $L_p$ norm. The nonlinearity, $\xi$, is introduced via a leaky rectified linear unit (leakyReLU) \cite{maas2013rectifier}.  

\begin{figure}[ht]
\centering
\includegraphics[width=0.96\textwidth]{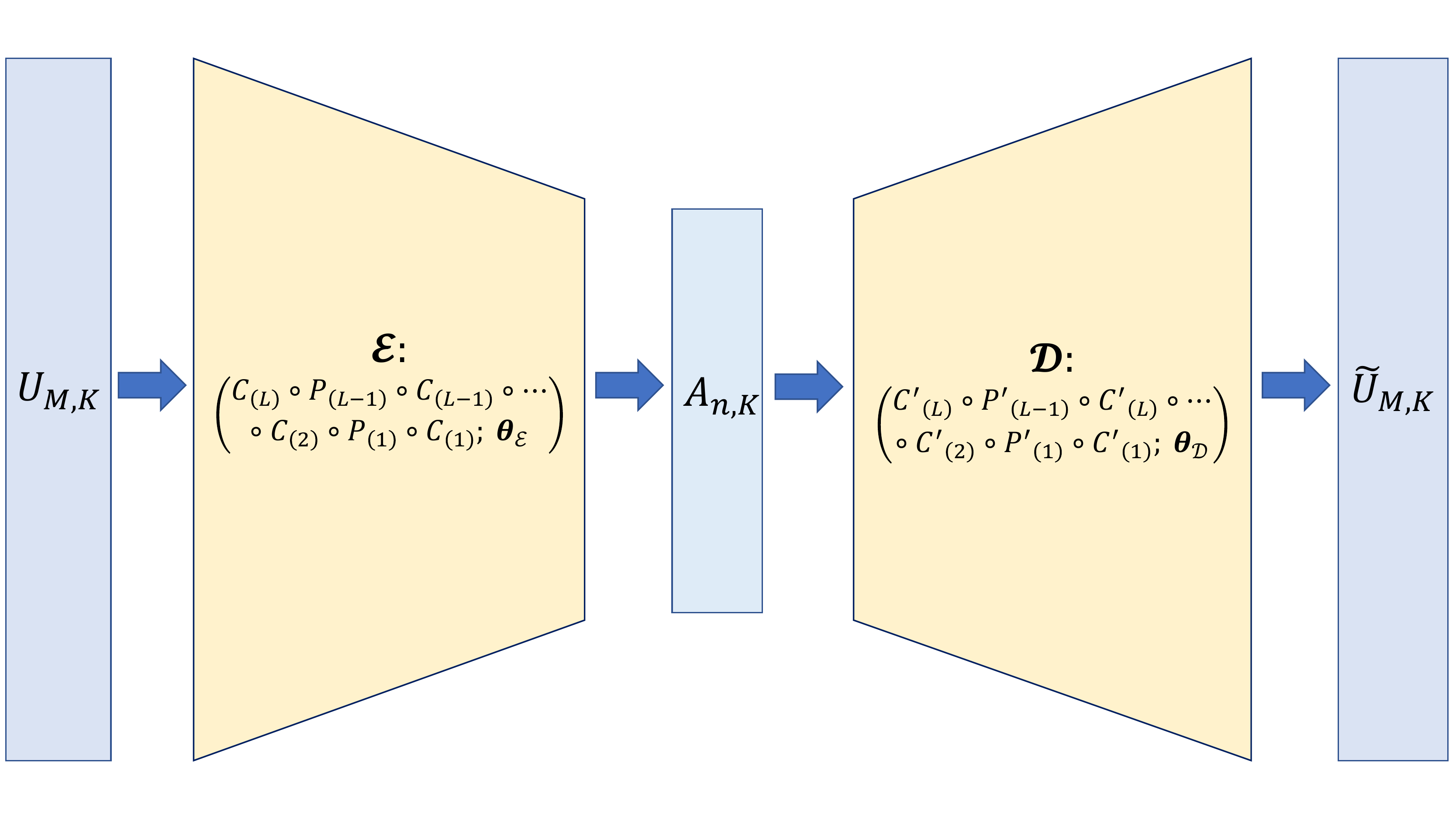}
\caption{An illustration of convolutional autoencoder for learning low-dimensional representation $A_{n,K}$, of high-dimensional data $U_{M,K}$}\label{fig:CAN-ROM}
\end{figure}

The convolutional encoder is a composition of $L$ convolutional layers and $L-1$ interleaved pooling layers, and has a general hierarchical representation,
\begin{equation}
    \label{eq:CNN_enc}
    \mathcal{E}(\bm{\mathrm{U}}) = \left(C_{(L)} \circ P_{(L-1)} \circ C_{(L-1)} \circ \ldots \circ C_{(2)} \circ P_{(1)} \circ C_{(1)};\bm{\theta}_\mathcal{E}\right)(\bm{\mathrm{U}}),
\end{equation}
where $\bm{\theta}_\mathcal{E} = \{\Lambda^{(1)}, \ldots,  \Lambda^{(2 L - 1)}\}$ is the set of all network parameters (all the kernel coefficients) of the convolutional encoder. The model is considered deep when it consists of multiple convolutional layers. For the present purpose we assume convolutional neural networks with three convolutional layers, which still classify as a deep model according to popular consensus. The convolutional decoder is obtained similarly, based on the output of the convolutional encoder,
\begin{equation}
    \label{eq:CNN_dec}
    \mathcal{D}(\bm{a}) = \left(C'_{(L)} \circ P'_{(L-1)} \circ C'_{(L-1)} \circ \ldots \circ C'_{(2)} \circ P'_{(1)} \circ C'_{(1)};\bm{\theta}_\mathcal{D}\right)(\bm{a}),
\end{equation}
where ${C'}(\bm{a}) = \xi\left({\kappa'}_\Lambda(\bm{a})\right)$ represent a transpose convolution layer and ${\kappa'}_\Lambda(\bm{a})$ represent a transpose convolution operation. Similarly, $P'$ represents an upsampling layer. Fig. \ref{fig:CAN-ROM} shows the dimensional reduction technique for the CAE. Our dimension-reduction objective is to find the lowest-dimensional subspace of dimension $n$ such that $\tilde{\bm{\mathrm{U}}}_{M,K} \approx \bm{\mathrm{U}}_{M,K}$.

Apart from the nonlinear compositional map generated by the CAE, certain geometric priors rely on the convolution and pooling operations \cite{bronstein2017geometric}, which help the network in both generalization and dimension reduction. For a translation operator
\begin{equation}
    \mathcal{T}_v U(x) = U(x-v), \quad x,v \in \Omega
\end{equation}
acting on functions $U \in L^2 \left(\Omega \right)$, the composition of convolutional layers endows translational equivariance, $y\left(\mathcal{T}_v U \right) =\mathcal{T}_v\left(y(U)\right)$, to the convolutional encoder and decoder. This can help in learning tasks like motion detection of dynamical systems and identification of coherent features over the domain. On the other hand, having pooling layers interleaved in convolutional layers endows translational invariance, $y\left(\mathcal{T}_v U \right) =y(U)$, to the convolutional encoder and decoder. Such translational invariance priors enable generalized classification of dynamical systems translated over time. 

Similarly, a local deformation $\mathcal{L}_\tau$ acting on $L^2 \left(\Omega \right)$ for a smooth vector field $\tau:\Omega \rightarrow \Omega$ can be defined as,
\begin{equation}
    \mathcal{L}_\tau U(x) = U\left(x - \tau(x)\right).
\end{equation}
Translation equivariance implies $\vert y\left(\mathcal{L}_\tau U\right) - \mathcal{L}_\tau y(U) \vert \approx \Vert \nabla \tau\Vert$ where $\Vert \nabla \tau\Vert$ measures the smoothness of a given deformation field. On the other hand, translational invariance implies $\vert y\left(\mathcal{L}_\tau U\right) - y(U) \vert \approx \Vert \nabla \tau\Vert$. Thus, both the translational equivariance and invariance priors imply that the prediction of the convolutional encoder and decoder networks does not change much even if the input is slightly deformed. Furthermore, the invariance to deformations can be interpreted from the perspective of a dynamical system as follows. The high-dimensional multi-scale interaction of dynamics can be separated into localized behavior according to their scales via downsampling in the hierarchical convolutional encoder. Such downsampling leads to a progressive reduction in the spatial resolution. Eventually, the most relevant scales of the dynamics are learned on a much lower-dimensional manifold via heuristic techniques. We will demonstrate the role of the geometric priors of the CAE in learning a low-dimensional representation later in the results section. We will also demonstrate and interpret in the results section how the geometric priors enable the CAE to generalize and not memorize.

\subsubsection{Projection-based dimension reduction}
By projecting the high-dimensional solution to a low-dimensional subspace, projection-based models such as POD attempt to find the best approximation of the high-dimensional solution. The best approximation is considered as the low-dimensional subspace that minimizes the $L_2$ error norm between the true and approximate solutions. Like most POD applications, we consider a Galerkin (i.e., orthogonal) projection. Here we briefly present the POD employed via the method of snapshots. An elaborate explanation can be found in Ref. \cite{rowley2017model}. 

Let us consider a set of snapshots $\bm{\mathrm{U}}_{M,K}={\left[\bm{\mathrm{u}}_1,\bm{\mathrm{u}}_2,\ldots,\bm{\mathrm{u}}_K\right]}^T$ obtained at $K$ time intervals, where $\bm{\mathrm{u}} \in \mathbb{R}^M$. POD assumes that the snapshot matrix can be decomposed into a linear superposition of spatial basis functions, each of which is associated with temporal coefficients
\begin{equation}
    \label{eq:POD}
    \bm{\mathrm{u}}_{M,k}\left(\bm{\mathrm{x}},t_k\right) = \sum_{j=1}^{K} \bm{\mathrm{a}}_{j,k} \left(t_k\right) \bm{\mathrm{v}}_j \left(\bm{\mathrm{x}}\right), \quad k=1,2,\ldots,K.
\end{equation}
Here we assume that the rank of $\bm{\mathrm{U}}$ is $K$ and $M > K$. In order to obtain these bases, we perform a singular value decomposition,
\begin{equation}
    \label{eq:SVD}
    \bm{\mathrm{U}} = \bm{\mathrm{V}} \bm{\mathrm{\Sigma}} \bm{\mathrm{W}}^T = \sum_{j=1}^{K} \sigma_j \bm{\mathrm{v}}_j {\bm{\mathrm{w}}_j}^T,
\end{equation}
where $\bm{\mathrm{V}}^T \bm{\mathrm{V}} = \bm{\mathrm{W}}^T \bm{\mathrm{W}} =\bm{\mathrm{I}}$. Thus, $\bm{\mathrm{V}}, \bm{\mathrm{W}} \in {\mathbb{R}}^{M\times K}$ and $\bm{\mathrm{\Sigma}}\in {\mathbb{R}}^{K\times K}$. The singular values are obtained from
\begin{equation}
    \label{eq:POD_sv}
    \bm{\mathrm{U}}^T \bm{\mathrm{U}} \bm{\mathrm{w}}_j = {\sigma_j}^2 \bm{\mathrm{w}}_j, \quad j=1,2,\ldots,K.
\end{equation}
Practical computation of ${\sigma_j}^2$ is performed via an eigenvalue analysis of Eq. (\ref{eq:POD_sv}) to obtain the POD modes as,
\begin{equation}
    \label{eq:POD_modes}
    \bm{\mathrm{v}}_j = \bm{\mathrm{U}} \bm{\mathrm{w}}_j/\sigma_j, \quad j=1,2,\ldots,K.
\end{equation}
For efficient model reduction, we intend to obtain the smallest set of $n$ modes 
\begin{equation}
    \begin{split}
        \label{eq:POD_reconstruct}
        \tilde{\bm{\mathrm{u}}}_{M,k}\left(\bm{\mathrm{x}},t_k\right) =& \sum_{j=1}^{n} \bm{\mathrm{a}}_{j,k} \left(t_k\right) \bm{\mathrm{v}}_j \left(\bm{\mathrm{x}}\right), \quad k=1,2,\ldots,K,\\
        \implies \tilde{\bm{\mathrm{U}}}_{M,K}\left(\bm{\mathrm{x}},t_K\right) &= \bm{\mathrm{V}}_n \left(\bm{\mathrm{x}}\right) \bm{\mathrm{A}}_{n,K} \left(t_K\right),
    \end{split}
\end{equation}
such that $\tilde{\bm{\mathrm{U}}}_{M,K} \approx \bm{\mathrm{U}}_{M,K}$.

POD is a widely used approach for approximating high-dimensional solutions to various physical problems. However, its efficiency is reduced when approximating solutions to hyperbolic partial differential equations and convection-dominated problems. The worst-case error of approximating high-dimensional solutions via projection to a lower-dimensional subspace of dimension $n$ is defined as the Kolmogorov $n$-width. It has been shown that the Kolmogorov $n$-width decays at a sub-exponential rate with $n$ even for linear transport problems once we consider non-uniform or discontinuous initial conditions \cite{greif2019decay}.
%\begin{figure}[ht]
%\centering
%\includegraphics[width=0.96\textwidth]{figures/Fig1.jpg}
%\caption{An illustration of proper orthogonal decomposition for learning low-dimensional representation $A_{n,K}$ from high-dimensional data $U_{M,K}$}
%\label{fig:POD-ROM}
%\end{figure}

\subsection{Learning system evolution}
The data-driven learning of the system evolution presented in Eq. (\ref{eq:data-driven_learning}) is posed as a sequence-to-sequence learning problem. To achieve this, we employ several deeply stacked LSTM networks. LSTMs are gated recurrent neural networks routinely used for accurately learning sequences with a long-term data dependency. The gating mechanism of LSTMs provides them invariance to time warping. Thus, they are significantly less affected by vanishing gradients compared to non-gated recurrent neural networks.

\begin{figure}[ht]
\centering
    \begin{subfigure}[b]{.45\textwidth}
        \centering
        \includegraphics[width=\textwidth]{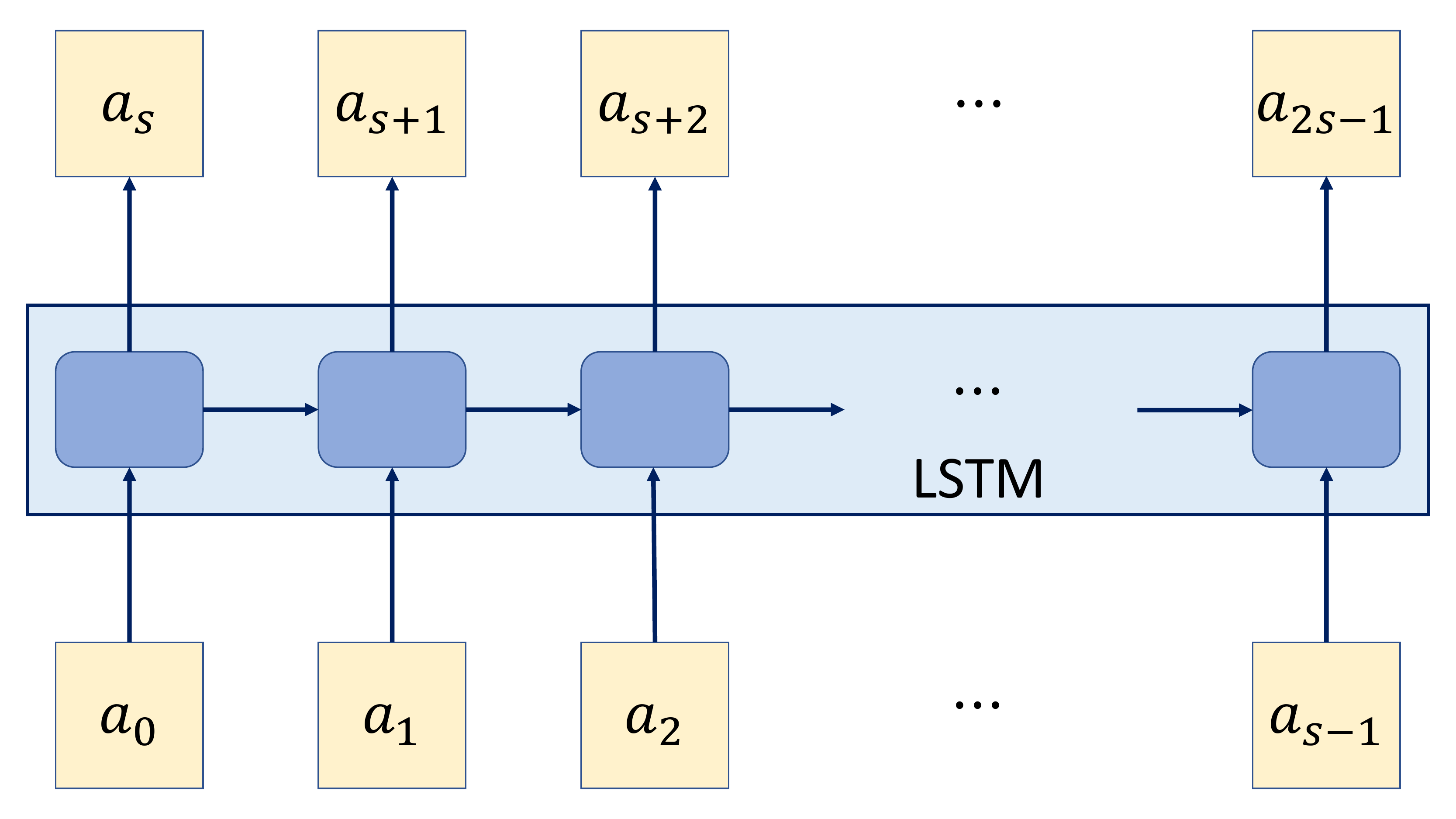}
        \caption{Standard LSTM}
    \end{subfigure}
    \hfill
    \begin{subfigure}[b]{.45\textwidth}
    \centering
    \includegraphics[width=\textwidth]{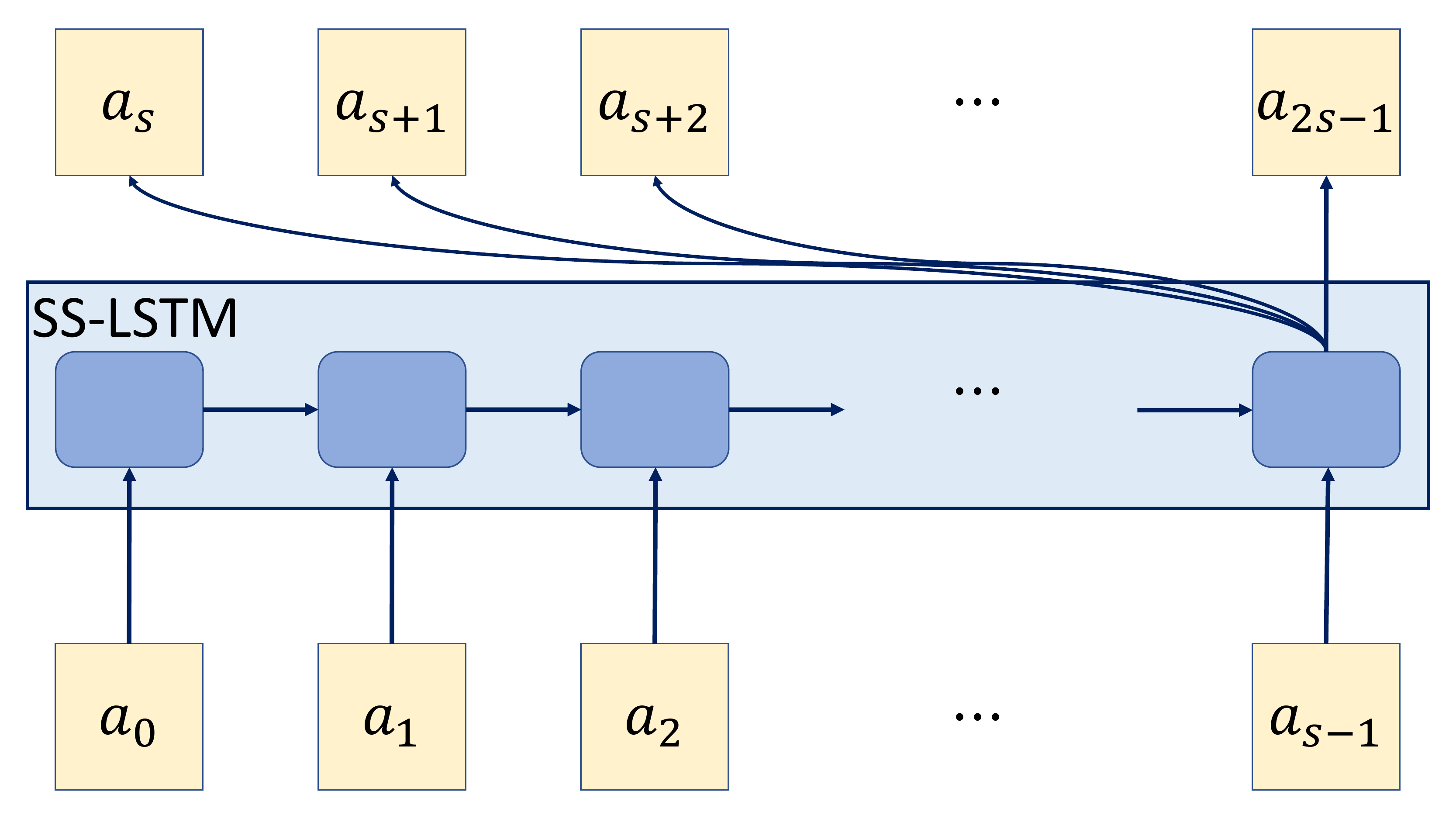}
         \caption{SS-LSTM}
    \end{subfigure}\\
    \begin{subfigure}[b]{.45\textwidth}
    \centering
    \includegraphics[width=\textwidth]{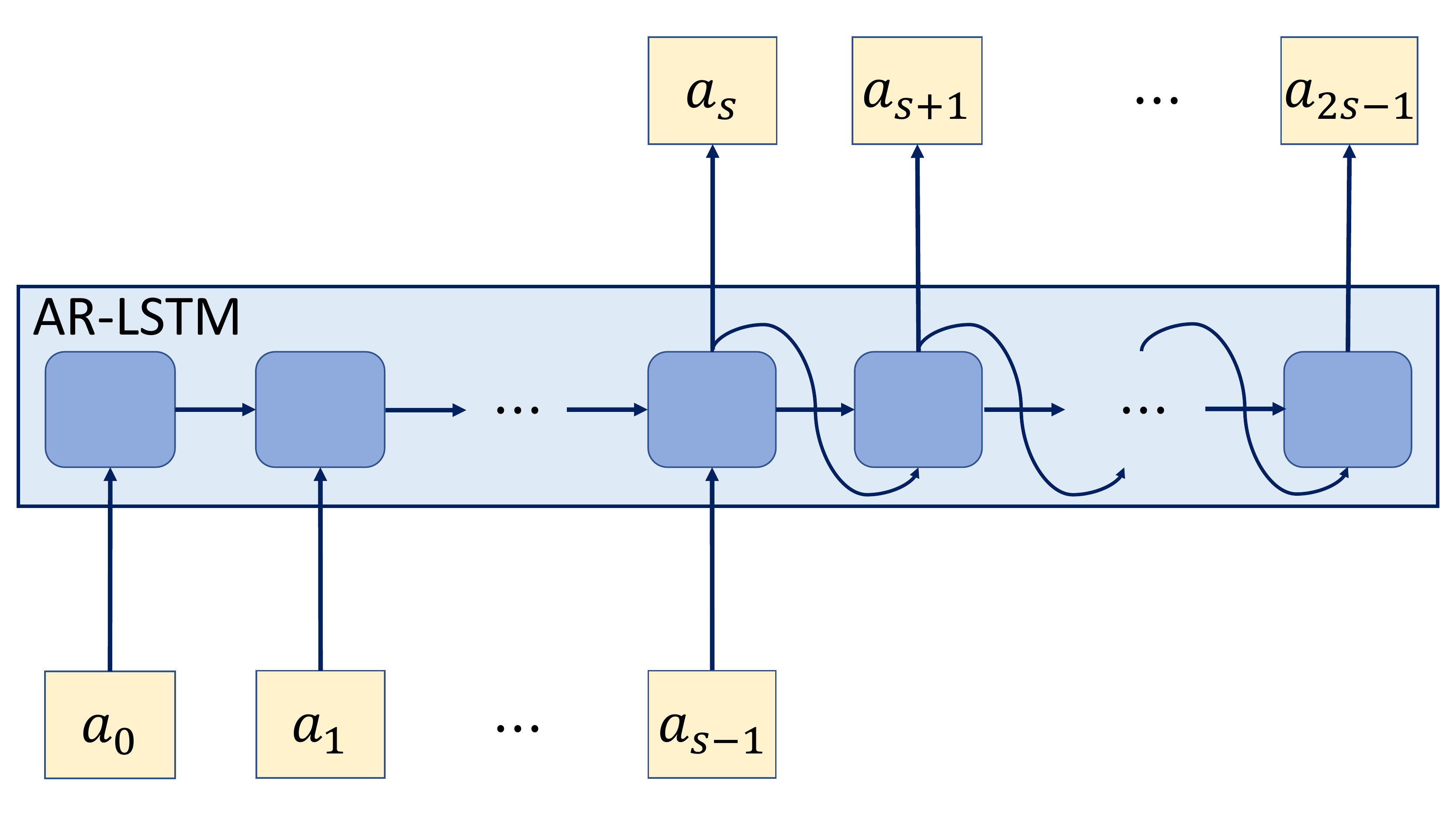}
         \caption{AR-LSTM}
    \end{subfigure}
\caption{Learning mechanism of various stacked LSTM architectures: (a) Standard LSTM, (b) SS-LSTM, and (c) AR-LSTM}
\label{fig:LSTM_architectures}
\end{figure}

A single LSTM cell consists of the input gate, the output gate, and the forget gate. The cell input, the cell state, and the cell output are denoted by $\bm{a}$, $\bm{c}$ and $\bm{h}$, respectively. The cell output is then passed to a fully connected layer with a linear activation to keep the input and output 
%($\bm{y}$)%
dimensions consistent. The operation of the LSTM cell can be explained via the following equations.
\begin{equation}
    \begin{split}
        \label{eq:LSTM_cell}
        \bm{f}_t =& \sigma \left(\bm{W}_f\cdot\left[\bm{h}_{t-1},\bm{a}_t \right]+\bm{b}_f\right),\\
        \bm{i}_t =& \sigma \left(\bm{W}_i\cdot\left[\bm{h}_{t-1},\bm{a}_t \right]+\bm{b}_i\right),\\
        \tilde{\bm{c}}_t =& \tanh \left(\bm{W}_c\cdot\left[\bm{h}_{t-1},\bm{a}_t \right]+\bm{b}_c\right),\\
        \bm{c}_t =& \bm{f}_t * \bm{c}_{t-1} + \bm{i}_t * \tilde{\bm{c}}_t,\\
        \bm{o}_t =& \sigma \left(\bm{W}_o\cdot\left[\bm{h}_{t-1},\bm{a}_t \right]+\bm{b}_o\right),\\
        \bm{h}_t =& \bm{o}_t * \tanh\left(\bm{c}_t\right),
    \end{split}
\end{equation}
where $\bm{i}$, $\bm{f}$ and $\tilde{\bm{c}}$ represent the input gate, the forget gate and the updated cell state, respectively. $\bm{W}$ and $\bm{b}$ represent the weights and biases for each of the gates, respectively. An illustration of the LSTM cell structure can be obtained in Ref. \cite{bukka2021assessment}. %On successful learning, we assume $\bm{y}_t\approx\bm{a}_{t+1}$. 
LSTM cells can be stacked together and their internal states can be connected to establish long-time dependency between the observables in a sequence. The gating mechanism of LSTMs ensures that correct values of the LSTM internal states connecting the observables are learned during the backpropagation phase of the training process. A plain stacked LSTM network with $s$-member sequence is shown in Fig. \ref{fig:LSTM_architectures} (a).

Here we are interested in multi-step sequence-to-sequence prediction, i.e. based on a sequence of observables, we will predict the next sequence of observables. The predicted sequence of observables can then act as input in an iterative manner to predict the evolution of the dynamical system over a long time horizon. The operation of plain LSTM with stacked LSTM cells can be mathematically represented as,
\begin{equation}
    \label{eq:plain_lstm_op}
    \begin{split}
    \bm{a}_s =& \mathcal{P}\left(\left[\bm{c}_0(\bm{a}_0),\bm{h}_0(\bm{a}_0)\right] ; \bm{\theta}_\mathcal{P} \right) \left(\bm{a}_0\right), \\
    \bm{a}_{s+1} =& \mathcal{P}\left(\left[\bm{c}_1(\bm{a}_1),\bm{h}_1(\bm{a}_1)\right] \circ \left[\bm{c}_0(\bm{a}_0),\bm{h}_0(\bm{a}_0)\right]; \bm{\theta}_\mathcal{P} \right) \left(\left[\bm{a}_1, \bm{a}_0\right]\right), \\
    \vdots &\\
    \bm{a}_{2s-1} =& \mathcal{P}\left(\left[\bm{c}_{s-1}(\bm{a}_{s-1}),\bm{h}_{s-1}(\bm{a}_{s-1})\right] \circ \ldots \circ \left[\bm{c}_0(\bm{a}_0),\bm{h}_0(\bm{a}_0)\right]; \bm{\theta}_\mathcal{P} \right) \left(\left[\bm{a}_{s-1}, \ldots , \bm{a}_0\right]\right),
    \end{split}
\end{equation}
where $\mathcal{P}$ represents the LSTM propagator and $\bm{\theta}_\mathcal{P}$ represents its trainable parameters. In this regard, we also explore the autoregressive LSTM (AR-LSTM) and the single-shot LSTM (SS-LSTM) architectures (Figs. \ref{fig:LSTM_architectures} (c) and (b), respectively), whose architectures resemble autoregressive models. For these autoregressive models, the prediction of the observable at any given time step depends on previous $s$ time steps via the composition of LSTM internal states ($\bm{c}$ and $\bm{h}$). The operation of the AR-LSTM can be mathematically represented as,
\begin{multline}
    \label{eq:ar_lstm_op}
    \bm{a}_k = \\ \mathcal{P}\left(\left[\bm{c}_{k-1}(\bm{a}_{k-1}),\bm{h}_{k-1}(\bm{a}_{k-1})\right] \circ \ldots \circ \left[\bm{c}_{k-s}(\bm{a}_{k-s}),\bm{h}_{k-s}(\bm{a}_{k-s})\right]; \bm{\theta}_\mathcal{P} \right) \left(\left[\bm{a}_{k-1}, \ldots , \bm{a}_{k-s}\right]\right),
\end{multline}
$k=s,s+1, \ldots, 2s-1$. Similarly, the operation of the SS-LSTM where the complete output sequence is learnt in a single shot from the input sequence, can be mathematically represented as
\begin{multline}
    \label{eq:ss_lstm_op}
    \left[\bm{a}_{2s-1}, \ldots , \bm{a}_s\right] = \\ \mathcal{P}\left(\beta\left(\left[\bm{c}_{s-1}(\bm{a}_{s-1}),\bm{h}_{s-1}(\bm{a}_{s-1})\right] \circ \ldots \circ \left[\bm{c}_0(\bm{a}_0),\bm{h}_0(\bm{a}_0)\right]\right); \bm{\theta}_\mathcal{P} \right) \left(\left[\bm{a}_{s-1}, \ldots , \bm{a}_0\right]\right).
\end{multline}
The operator $\beta$ ensures that appropriate maps are generated for each of the output sequence members from the composition of the input LSTM states.

The autoregressive modeling, where each sequence member relies on $s$ previous sequence members, is absent in the plain LSTM architecture (Eq. \ref{eq:plain_lstm_op}). This leads to poor generalizations in plain LSTM architectures, especially for iterative LSTM applications to predict system dynamics over a long time horizon \cite{bukka2021assessment,deo2022predicting}. Both these articles \cite{bukka2021assessment,deo2022predicting} prefer context-driven encoder-decoder variants of plain LSTM architectures, e.g., attention-mechanism in article \cite{deo2022predicting}. In essence, these variations in the plain LSTM architecture enable autoregressive modeling via dependency on previously predicted observables. This motivates us to employ 
the AR-LSTM and SS-LSTM architectures in the CRAN model for the present study.

\subsection{Convolutional recurrent autoencoder architecture}
\begin{figure}[ht!]
    \centering
    \includegraphics[width=1.0\linewidth]{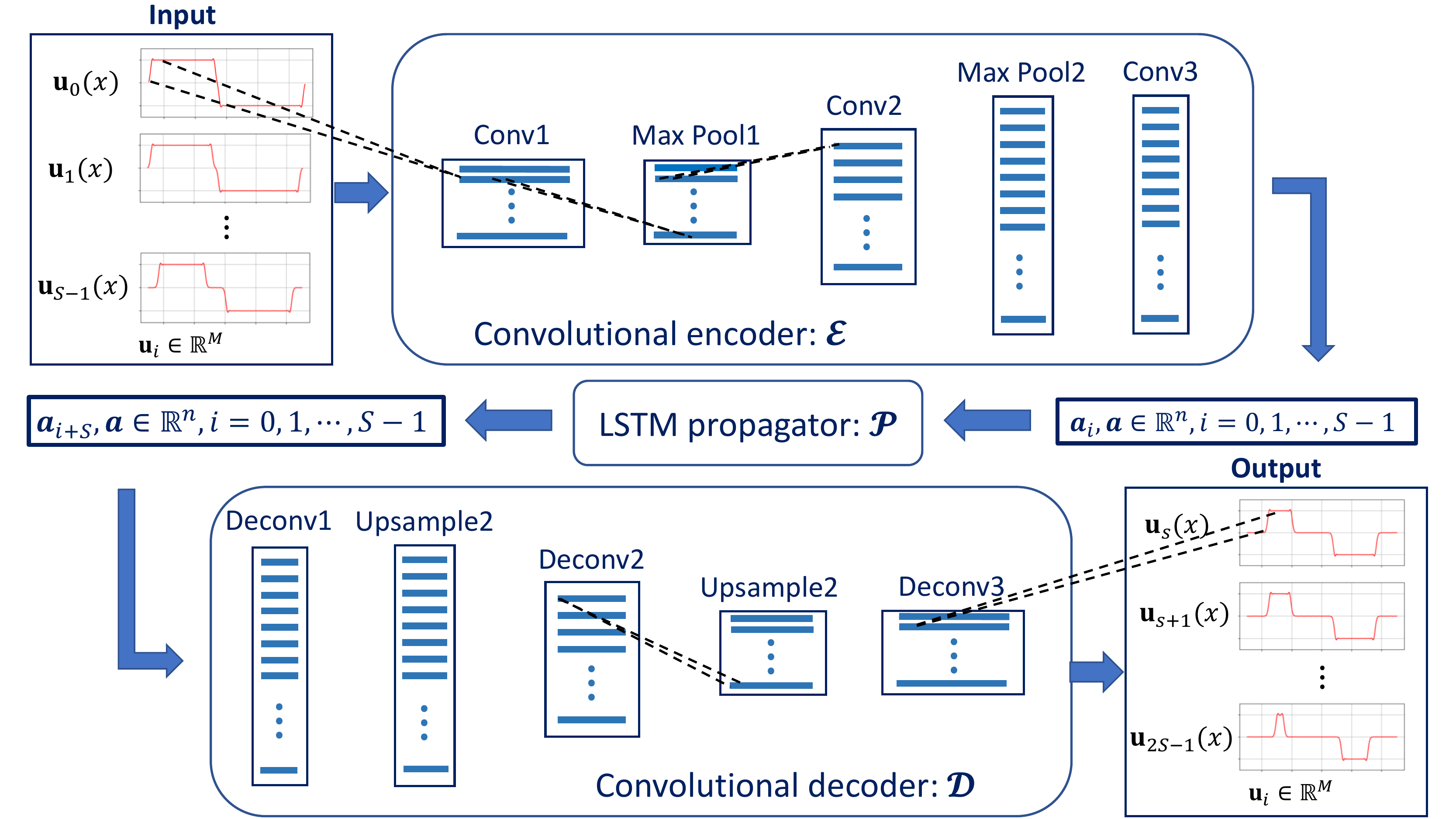}
    \caption{Illustration of the 1D CRAN for data-driven learning}
    \label{fig:1D-CRAN}
\end{figure}

The two separate data-driven learning tasks discussed previously are performed via the composite convolutional recurrent autoencoder network (CRAN) shown in Fig. \ref{fig:1D-CRAN}. Here we present a three-layer convolutional encoder $\mathcal{E}$ and decoder $\mathcal{D}$. The encoder and decoder presented here are generated with one-dimensional convolutional kernels, which operate on one-dimensional inputs. However, the convolution encoding and decoding can be performed in an analogous manner for higher-dimensional Euclidean data sets. The LSTM propagator $\mathcal{P}$ presented here is any general LSTM model. Thus, any one of the three LSTM architectures presented above can be included in the CRAN model.  

We have separated the data-driven learning task into a lower-dimensional spatial representation learning task and a system evolution learning task. The CAE (Fig. \ref{fig:CAN-ROM}) learns the low-dimensional representation of the high-dimensional physical data, and the LSTM network learns the system evolution. These are trained separately. On successful training of these individual components, the CRAN model is employed during the prediction phase in a sequence-to-sequence manner while operating on a low-dimensional subspace. For a sequence containing $S$ units, this can be mathematically represented as
\begin{equation}
    \label{eq:CRAN_prediction}
    \begin{split}
        \bm{\mathrm{A}}_{n,S_0}\left(t_{S_0};\phi \right) =& \mathcal{E}\left(\bm{\mathrm{U}}_{M,S_0}\left(\bm{\mathrm{x}}_M, t_{S_0};\phi \right);\bm{\theta}_\mathcal{E} \right),\\
        \bm{\mathrm{A}}_{n,S_1}\left(t_{S_1};\phi \right) =& \mathcal{P}\left(\bm{\mathrm{A}}_{n,S_0}\left(t_{S_0};\phi \right);\bm{\theta}_\mathcal{P}\right),\\
        \tilde{\bm{\mathrm{U}}}_{M,S_1}\left(\bm{\mathrm{x}}_M, t_{S_1};\phi \right) =&\mathcal{D}\left(\bm{\mathrm{A}}_{n,S_1}\left(t_{S_1};\phi \right);\bm{\theta}_\mathcal{D}\right),
    \end{split}
\end{equation}
where $\bm{\mathrm{U}}_{M,S_0}$, $\tilde{\bm{\mathrm{U}}}_{M,S_1} \in \mathbb{R}^{M\times S}$ and $\bm{\mathrm{A}}_{n,S_0}$, $\bm{\mathrm{A}}_{n,S_1} \in \mathbb{R}^{N\times S}$. Feeding back $\tilde{\bm{\mathrm{U}}}_{M,S_1}$ we can obtain $\tilde{\bm{\mathrm{U}}}_{M,S_2}$ as the output. Autoregressive application of CRAN over a large number of iterations $r$ can enable us to obtain $\tilde{\bm{\mathrm{U}}}=\left[\tilde{\bm{\mathrm{U}}}_{M,S_1}, \tilde{\bm{\mathrm{U}}}_{M,S_2}, \ldots, \tilde{\bm{\mathrm{U}}}_{M,S_r}\right], \tilde{\bm{\mathrm{U}}} \in \mathbb{R}^{M \times r S}$. Thus, starting with $S$ initial propagation steps, the autoregressive application of the CRAN can evolve the system to $r \times S$ propagation steps into the horizon. The accuracy of the prediction $\tilde{\bm{\mathrm{U}}}$ will indicate CRAN's data-driven learning capability. When the CRAN is applied for learning wave propagation, the observable ${\bm{\mathrm{U}}}$ can be replaced with the discrete acoustic pressure observable, ${\bm{\mathrm{p}}}$ (Eq. \ref{eq:datadrivenacouspres}) without any loss in generality.

\section{Test scenarios for CRAN-based wave propagation learning}
In the present study, we consider a one-dimensional homogeneous wave equation, such that $q=0$ in Eq. \ref{eq:2ndorder_wave_PDE}. We consider a fully reflecting boundary by applying pressure release boundary conditions,
\begin{equation}
    p\left(-\frac{L}{2},t\right) = p\left(\frac{L}{2},t\right) = 0.
\end{equation}
Finally, we employ a spatially varying discontinuous initial condition depending on spatial location $x_s$,
\begin{equation}
    \label{eq:discont_IC}
    p\left(x,0;x_s \right) = \begin{cases}
    1, &\quad \mathrm{if} \ x < x_s, \\
    -1, &\quad \mathrm{if} \ x \geq x_s,
    \end{cases}
    \quad x \in \left[-\frac{L}{2},\frac{L}{2}\right].
\end{equation}
For the aforesaid conditions, there is no closed-form solution to Eq. (\ref{eq:2ndorder_wave_PDE}). Thus, a Galerkin-finite element method was employed to obtain the pressure fluctuations. Temporal snapshots of the pressure fluctuations obtained at regular intervals will serve as the training data for the CAE and the LSTM networks. For the bounded domain, the interference pattern obtained is periodic in nature. Thus, training data obtained over a single period is sufficient to train the network. 

Here, we are also interested in the generalized learning and prediction capability of the data-driven model with variation in the location of the discontinuous initial condition (i.e., varying $x_s$ in Eq. \ref{eq:discont_IC}). Thus, temporal snapshots over one period of evolution for several spatially varying locations of the discontinuity were obtained to train the network. The generalized learning capacity of the CRAN model will be determined by its ability to predict solutions for any general location of discontinuity $x_s$ over the domain, which is not included in the training set. During the prediction phase, we employ autoregressive CRAN operation following Eq. \ref{eq:CRAN_prediction}, to compute one period of the evolution of the pressure fluctuation. It is important to note that wave propagation with spatially varying discontinuities can be considered a translationally equivariant phenomenon prior to reflection from any boundary. This can be easily observed when the general solution of Eq. \ref{eq:2ndorder_wave_PDE}, with the initial conditions Eq. \ref{eq:discont_IC}, is written as a superposition of the left and right traveling characteristics,
\begin{equation}
    \label{eq:Greens_function_1Dplanewaves}
    {p}(x,t;x_s) = \mathcal{F}_1 \left(t + \frac{x-x_s}{c_0}\right) + \mathcal{F}_2 \left(t - \frac{x-x_s}{c_0}\right).
\end{equation}
The translational equivariant geometric priors associated with CAEs have been discussed earlier. This further motivates the employment of convolutional neural networks in the CRAN for generalized learning and classification of spatially varying parametric wave propagation data. 

\section{\label{sec:6} Results}
In this section we will present the results obtained with the CRAN model for learning the wave propagation problem. The capability of CAEs to reduce the system on a low-dimensional manifold, and the learning of system evolution via various LSTM models, will be discussed. Subsequently, we will provide a physical interpretation of the dimension reduction and generalized learning of wave propagation phenomena via the CAEs. The CAE and the LSTM networks employed in this study were developed and trained with TensorFlow 2.5.0 \cite{tensorflow2015-whitepaper} libraries. The training was performed via ADAM optimizer \cite{kingma2014adam}, which minimizes the mean square error between the target and predicted training values. 

\begin{figure}[ht]
\centering
\includegraphics[width=0.49\linewidth]{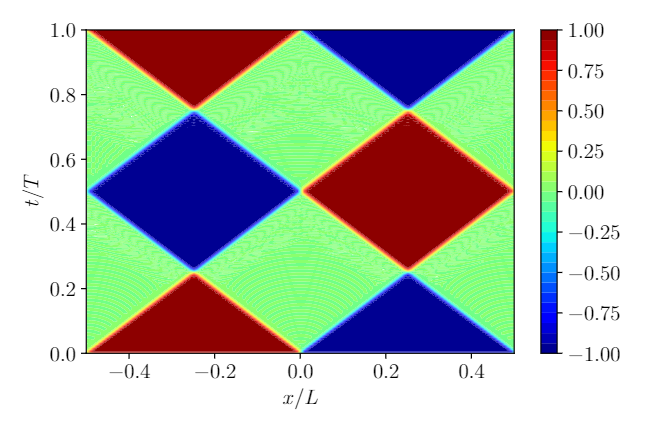}
\caption{Normalized pressure fluctuations for discontinuous initial condition obtained via finite element method: $x_s/L=0$}
\label{fig:1D_discont_soln_xs0}
\end{figure}

The ground truth for the second-order wave propagation with discontinuous initial conditions is computed via a finite element solver. The finite element method-based solution of the pressure fluctuations obtained for $x_s=0$ is presented in Fig. \ref{fig:1D_discont_soln_xs0}. The wave propagation solution is obtained over one complete period and consists of 512 temporal snapshots over uniform time intervals. The spatial data is uniformly discretized on $M$ points with $M=1025$. This solution will serve as the baseline case to compare the efficiency of POD and CAE for learning low-dimensional spatial representation. It will also serve as a baseline to compare the various LSTM architectures discussed earlier and select the most accurate and efficient network.

\subsection{Dimension reduction on low-dimensional manifold}
Discontinuities in the initial condition can pose a challenge in the dimension reduction of hyperbolic partial differential equation solutions via projection-based models.
A recent study \cite{greif2019decay} claims that for such initial conditions, the Kolmogorov $n$-width (worst-case error from projection on optimal linear subspaces) decays only as $\mathcal{O}\left( n^{1/2}\right)$. We numerically inspect such results with POD on a bounded domain and then compare them with the dimension reduction efficiency of the CAE. 

\begin{figure}[ht]
\centering
\includegraphics[width=0.49\textwidth]{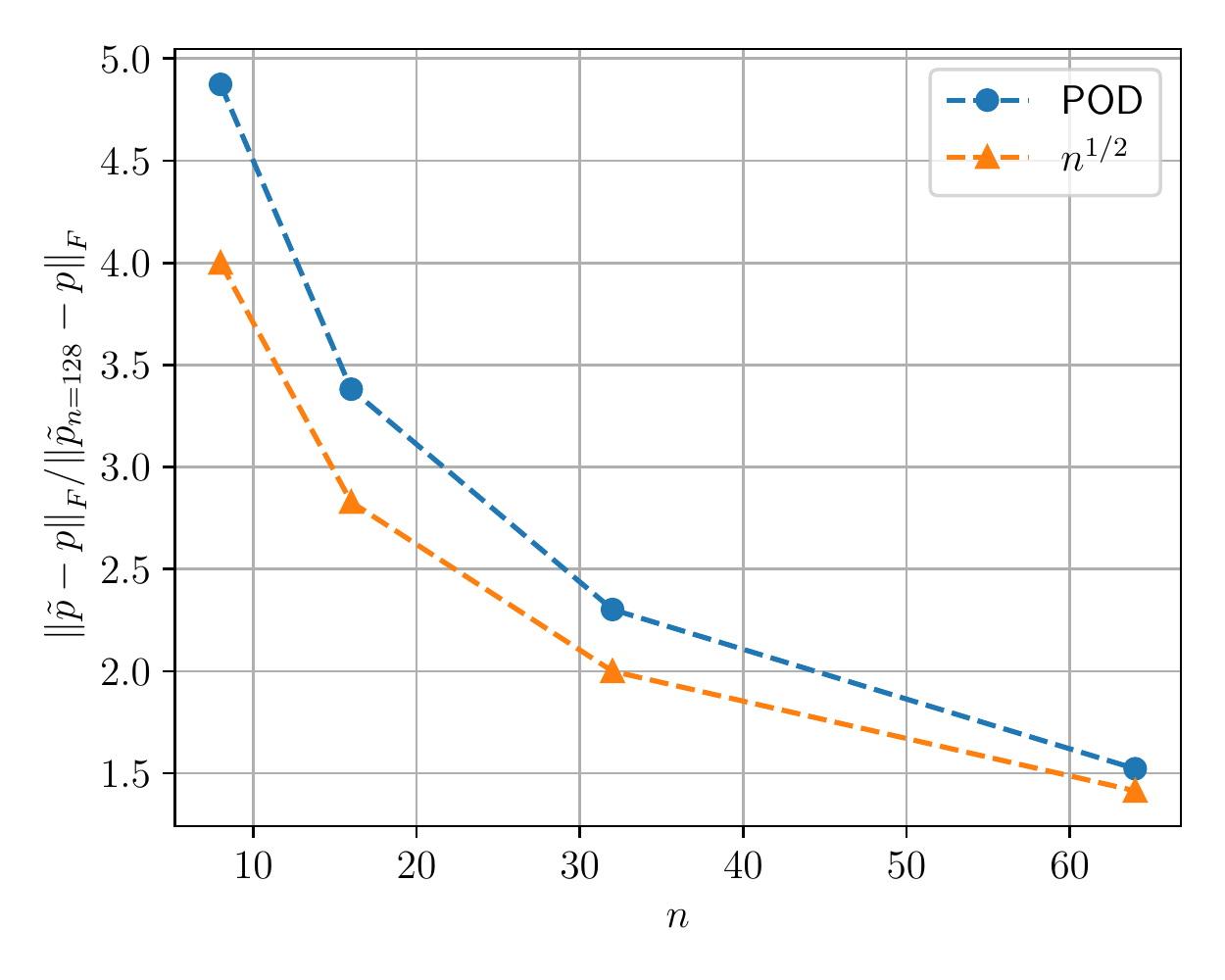}
\caption{Comparison of relative Frobenius error norms of POD models with $n=8,16,32,64$ relative to $n=128$, to theoretical decay rate of $d_n\left(\mathcal{M}\right)$}
\label{fig:comp_POD_theory}
\end{figure}

Fig. \ref{fig:comp_POD_theory} shows the relative Frobenius error norm of POD predictions for $n=8,16,32,64$ relative to POD predictions for $n=128$. Here $\tilde{p}$ and $p$ are predicted and target pressure fluctuations, respectively. We also show the theoretical error decay at $\mathcal{O}\left( n^{1/2}\right)$ to compare with the POD error decay. The results show that POD prediction error indeed decays at a sub-exponential rate indicating the inefficiency of the projection-based approach for reducing the high-dimensional physical system on a low-dimensional manifold. 

\begin{figure}[ht]
\centering
\includegraphics[width=0.49\linewidth]{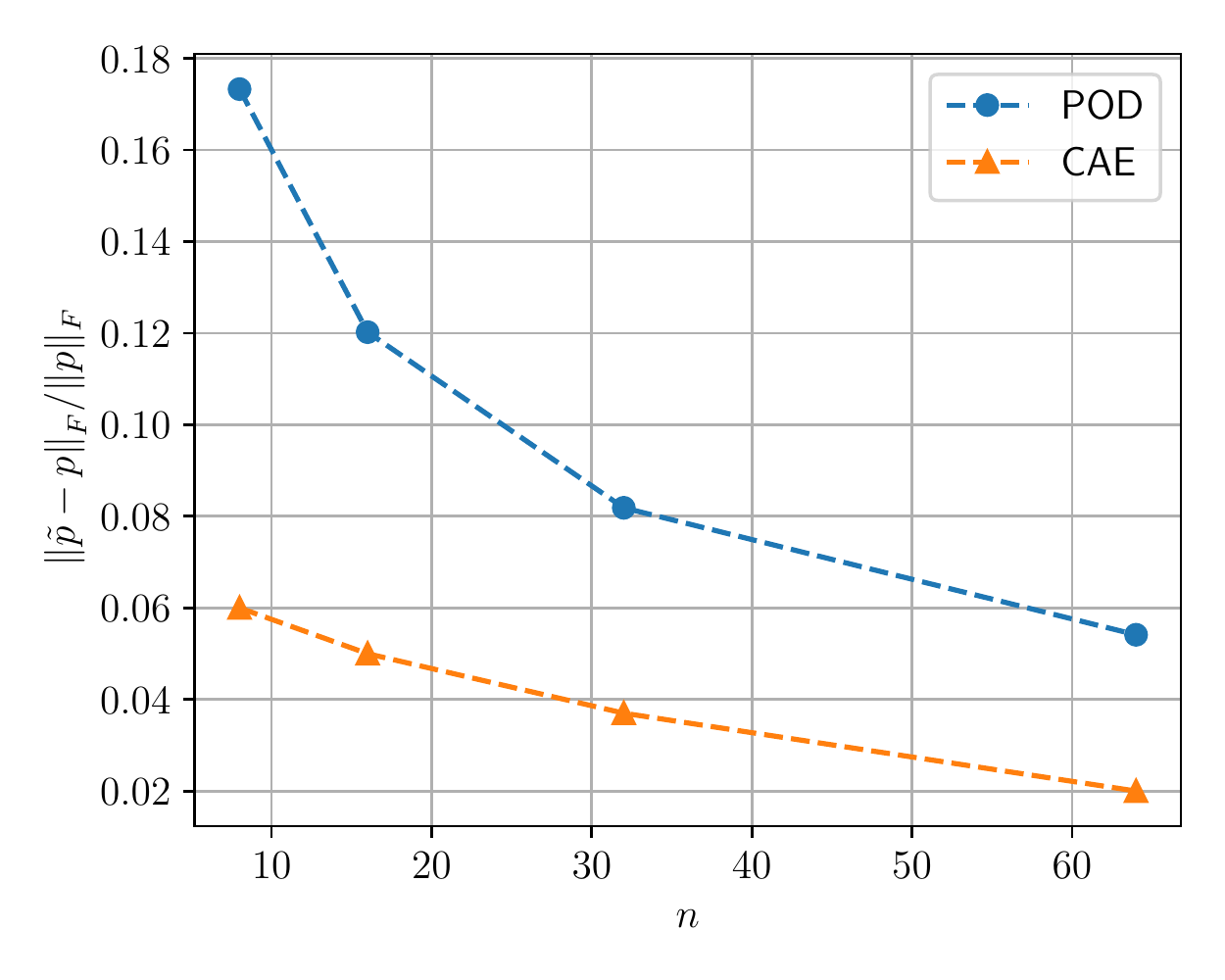}
\caption{Comparison of relative Frobenius error norms of various POD and CAE model predictions with $n=8,16,32,64$}
\label{fig:POD_CAN_relerr_comp}
\end{figure}

In Fig. \ref{fig:POD_CAN_relerr_comp} the relative Frobenius error norms computed with respect to the true solution of various POD models with $n=8,16,32,64$ are illustrated. The counterparts of such error norms were obtained from the CAE trained with the baseline result as both training input and target output. The CAE training errors are presented in Fig. \ref{fig:POD_CAN_relerr_comp} for $n=8,16,32,64$. The results show that even with 64 modes, the POD predictions incur 5.6\% error. However, the CAE predictions with only eight latent states show 6\% error, which reduces to 2\% when 64 latent states are considered. The results indicate the CAE can reduce the high-dimensional data almost three times more efficiently compared to the projection-based approach.

\begin{figure}[ht]
\centering
\includegraphics[width=0.49\linewidth]{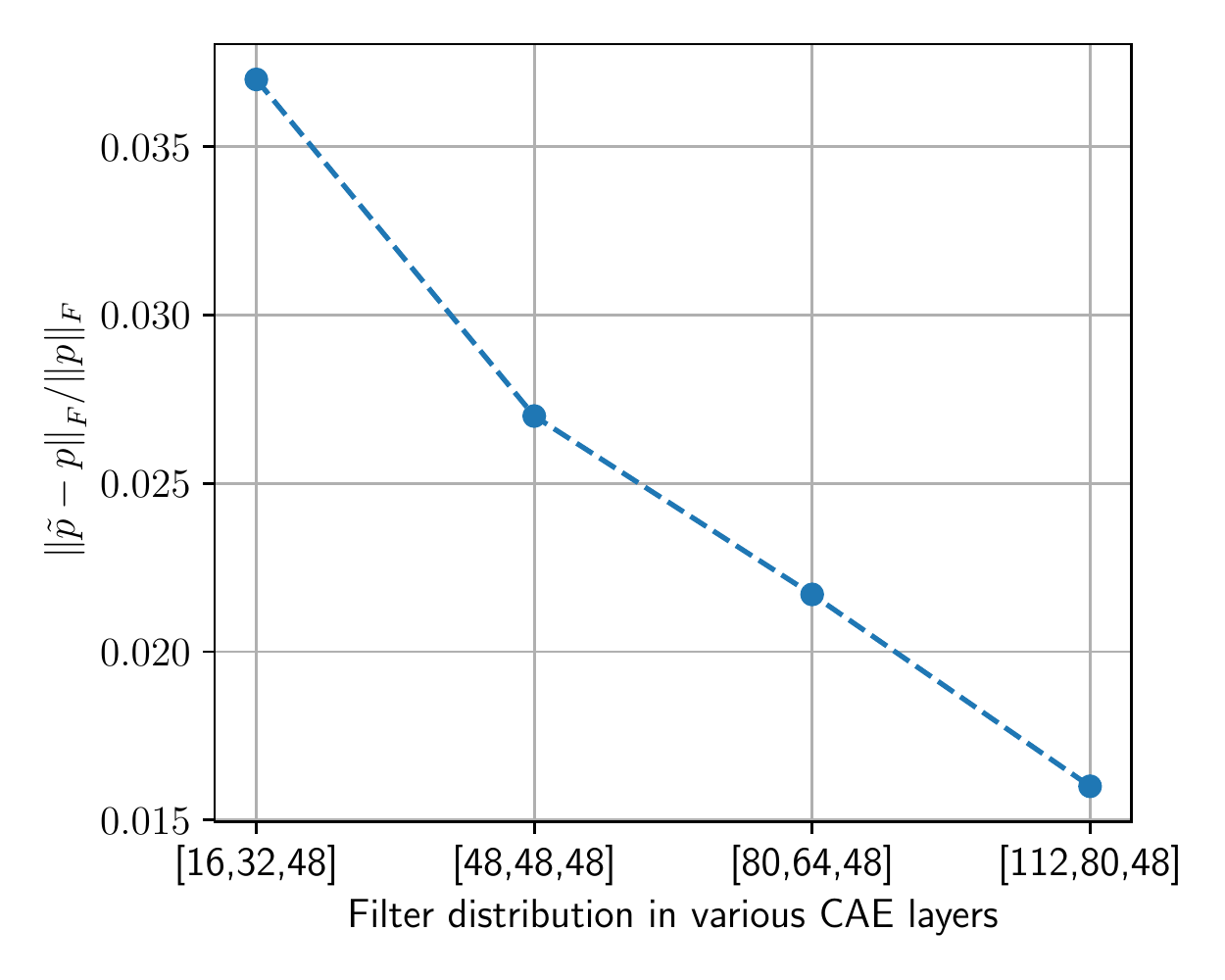}
\caption{Comparison of relative Frobenius error norms of various CAE models with $n=32$ and varying filter distribution in the first and second layers}
\label{fig:comp_CANls32varyingnf}
\end{figure}

As with any heuristic model, the CAE has several hyperparameters, which can be tuned to obtain a more accurate low-dimensional representation. One such hyperparameter is the filter distribution in various layers of the CAE. For the results presented in Fig. \ref{fig:POD_CAN_relerr_comp} the number of filters $n_f$ were 16, 32 and 48 in the first, second and third layers, respectively, of the CAE. However, as we can see from the relative error norms of the CAE with $n=32$ in Fig. \ref{fig:comp_CANls32varyingnf}, increasing the number of filters can significantly enhance the dimension reduction of the autoencoder model. In this comparison, the number of filters in the innermost layer is kept constant while the first and second layer filter distribution is changed. We will physically interpret the effect of filter distribution on the low-dimensional representation learning ability of the CAE later in the discussion.

\begin{figure}[ht]
\centering
    \begin{subfigure}[b]{.32\textwidth}
        \centering
        \includegraphics[width=\textwidth]{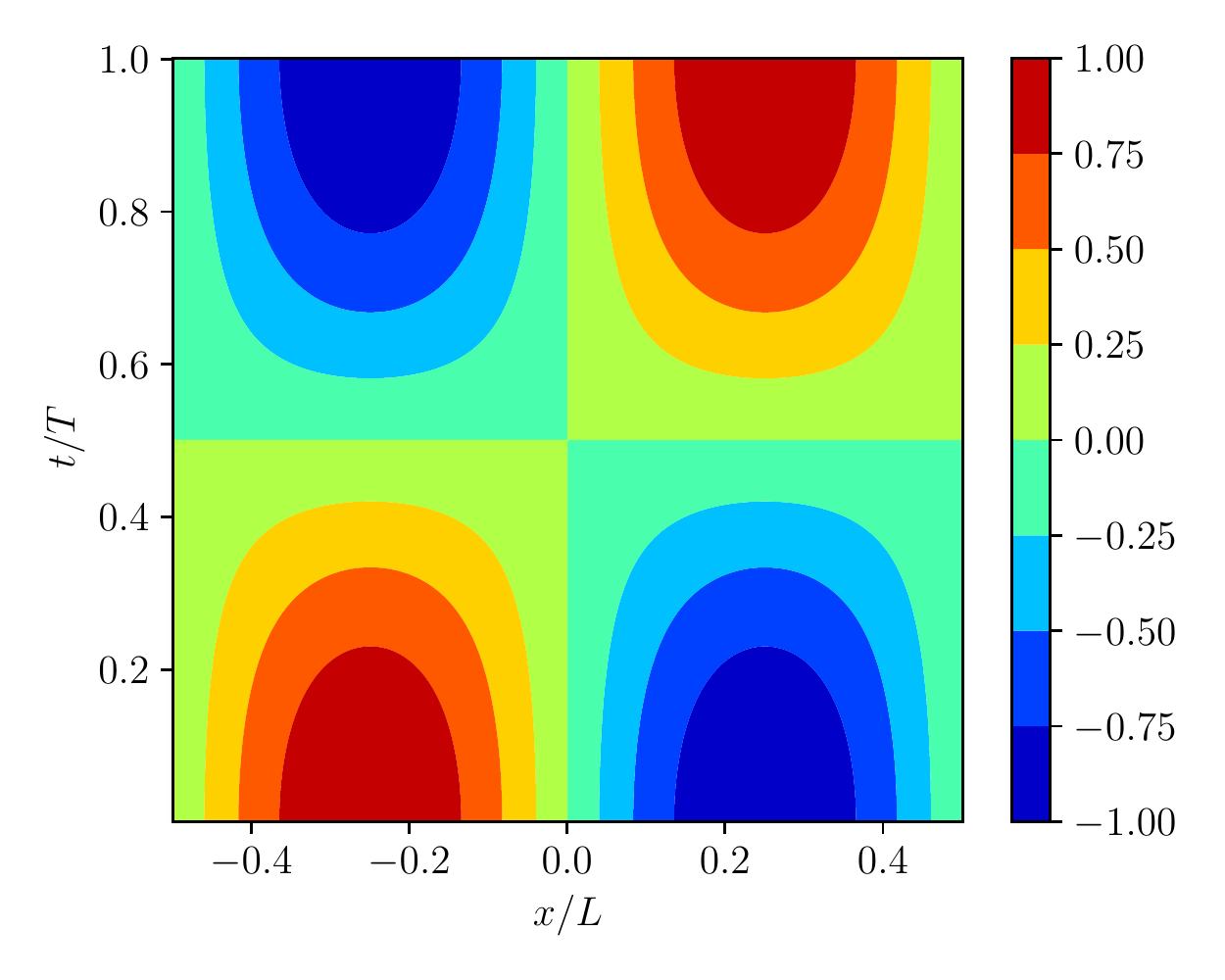}
        \caption{POD mode 1}
    \end{subfigure}
    \hfill
    \begin{subfigure}[b]{.32\textwidth}
    \centering
    \includegraphics[width=\textwidth]{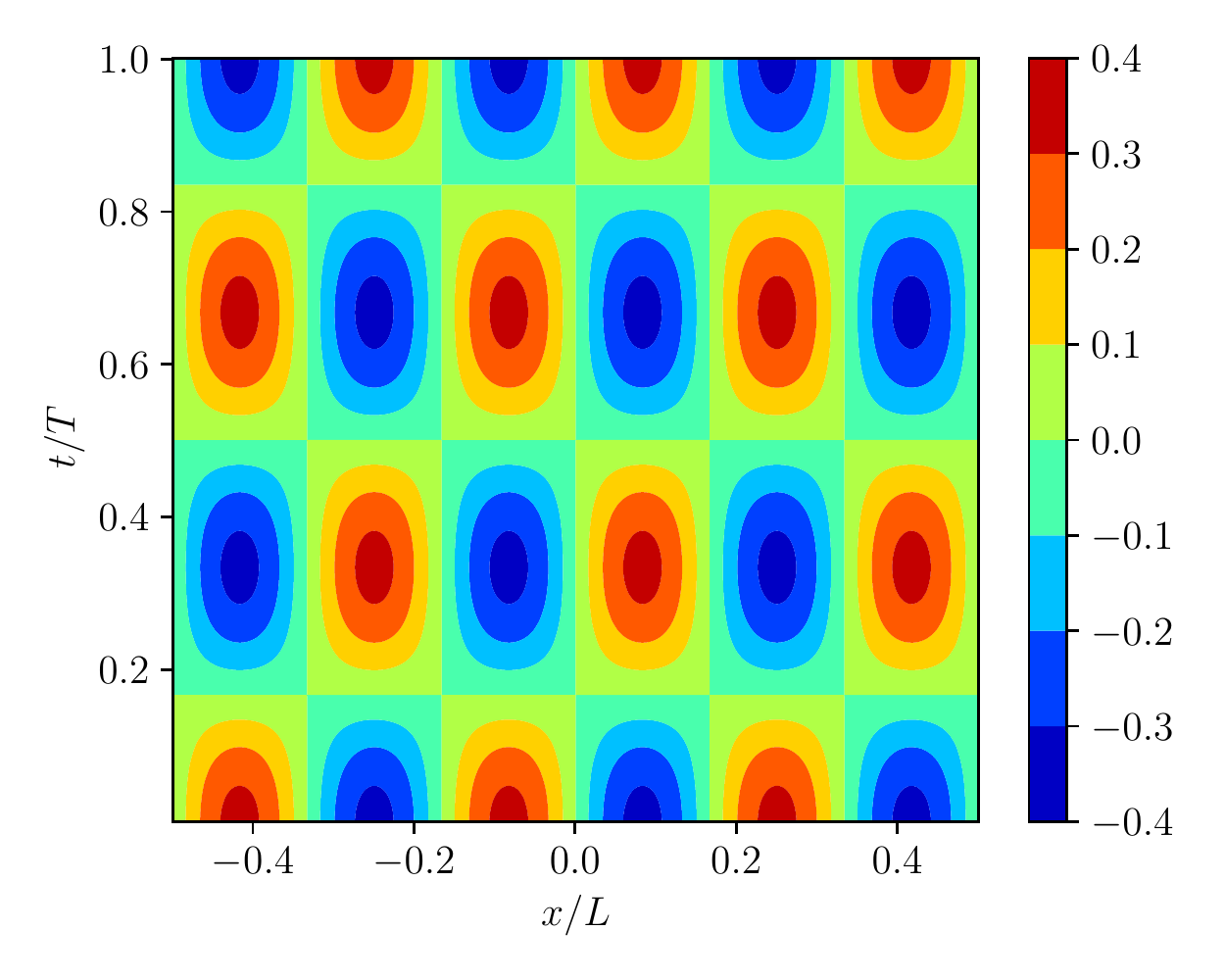}
         \caption{POD mode 2}
    \end{subfigure}
    \begin{subfigure}[b]{.32\textwidth}
    \centering
    \includegraphics[width=\textwidth]{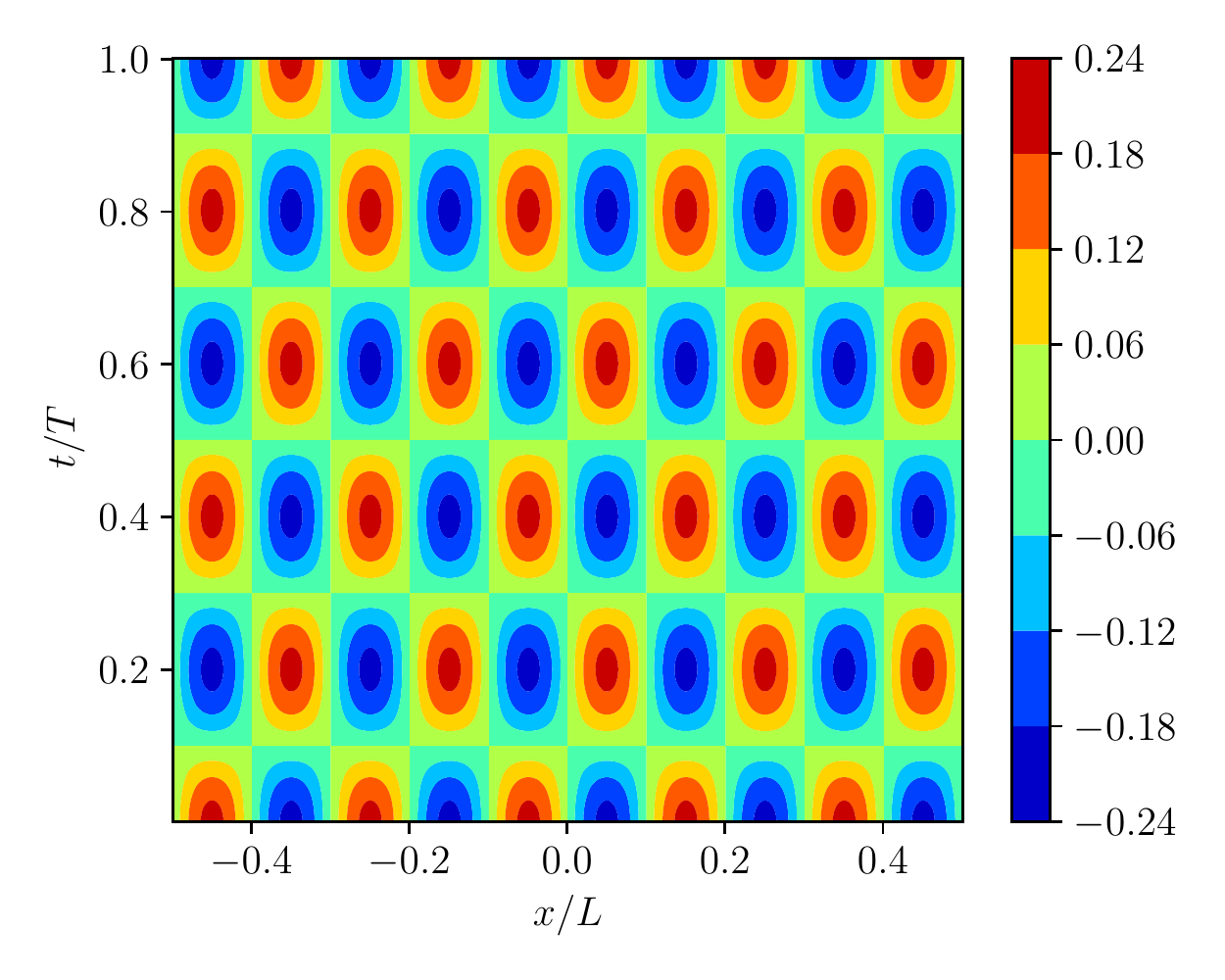}
         \caption{POD mode 3}
    \end{subfigure}
\caption{Temporal evolution of POD modes 1, 2 and 3}
\label{fig:evolvePODmodes1to3}
\end{figure}

To understand why the projection-based approach suffers from a slowly decaying Kolmogorov $n$-width, we inspect the evolution of POD modes 1-3 in Fig. \ref{fig:evolvePODmodes1to3}. It can be observed that the POD modes obtained via projecting the high-dimensional system onto the optimal linear subspace results in Fourier modes. However, such Fourier modes are quite restrictive in capturing discontinuities in hyperbolic partial differential equation solutions, where very little dissipation/diffusion exists. Thus, projection-based approaches are simply not efficient for dimension reduction for such physical problems. 
%\begin{figure}[ht]
%\centering
%\includegraphics[width=0.49\linewidth]{figures/PODmodeshapes1to3.pdf}
%\caption{The POD mode shapes 1, 2 and 3}
%\label{fig:PODmodeshapes1to3}
%\end{figure}

\begin{figure}[ht]
\centering
    \begin{subfigure}[b]{.32\textwidth}
        \centering
        \includegraphics[width=\textwidth]{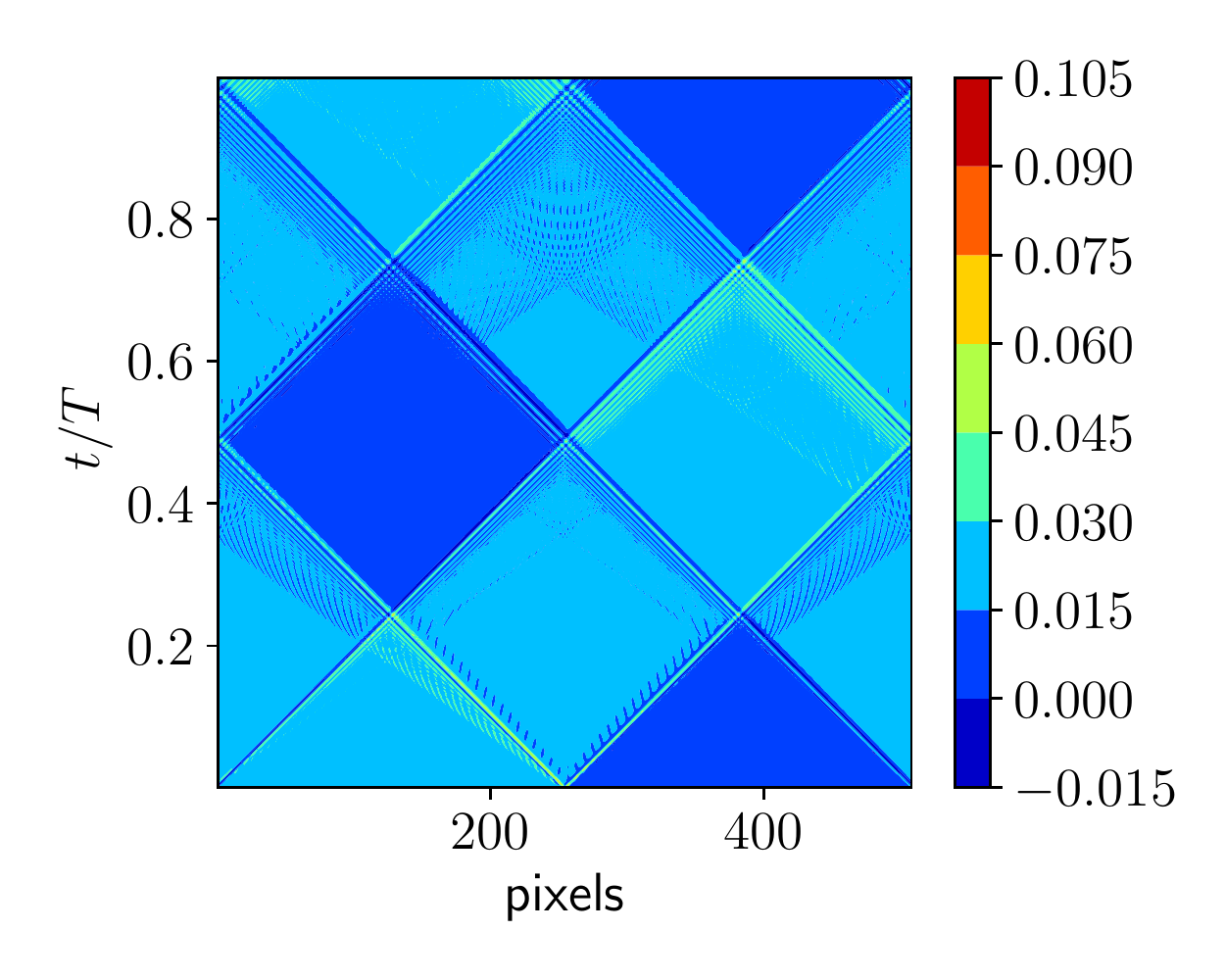}
        \caption{$f=3$}
    \end{subfigure}
    \hfill
    \begin{subfigure}[b]{.32\textwidth}
    \centering
    \includegraphics[width=\textwidth]{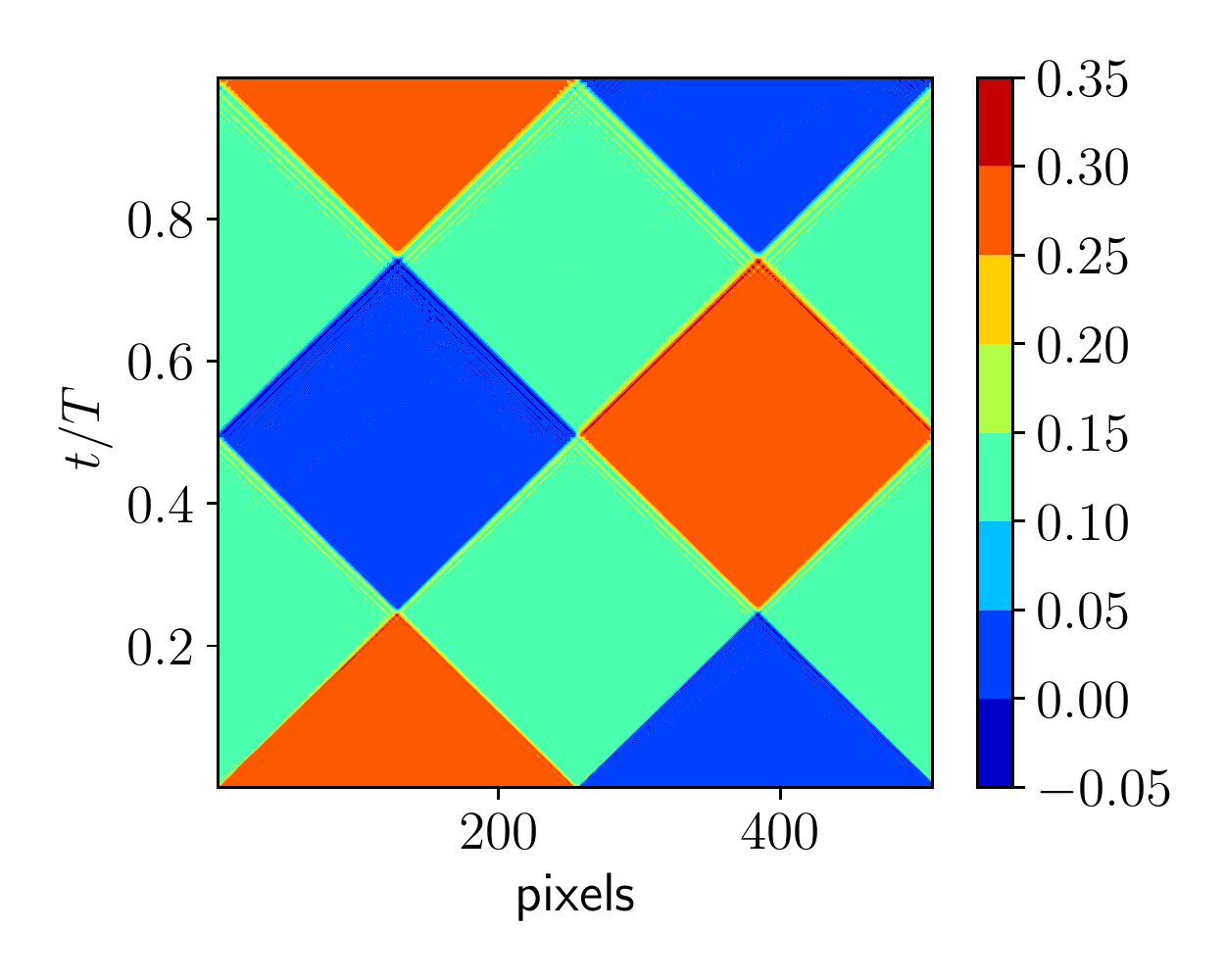}
         \caption{$f=9$}
    \end{subfigure}
    \begin{subfigure}[b]{.32\textwidth}
    \centering
    \includegraphics[width=\textwidth]{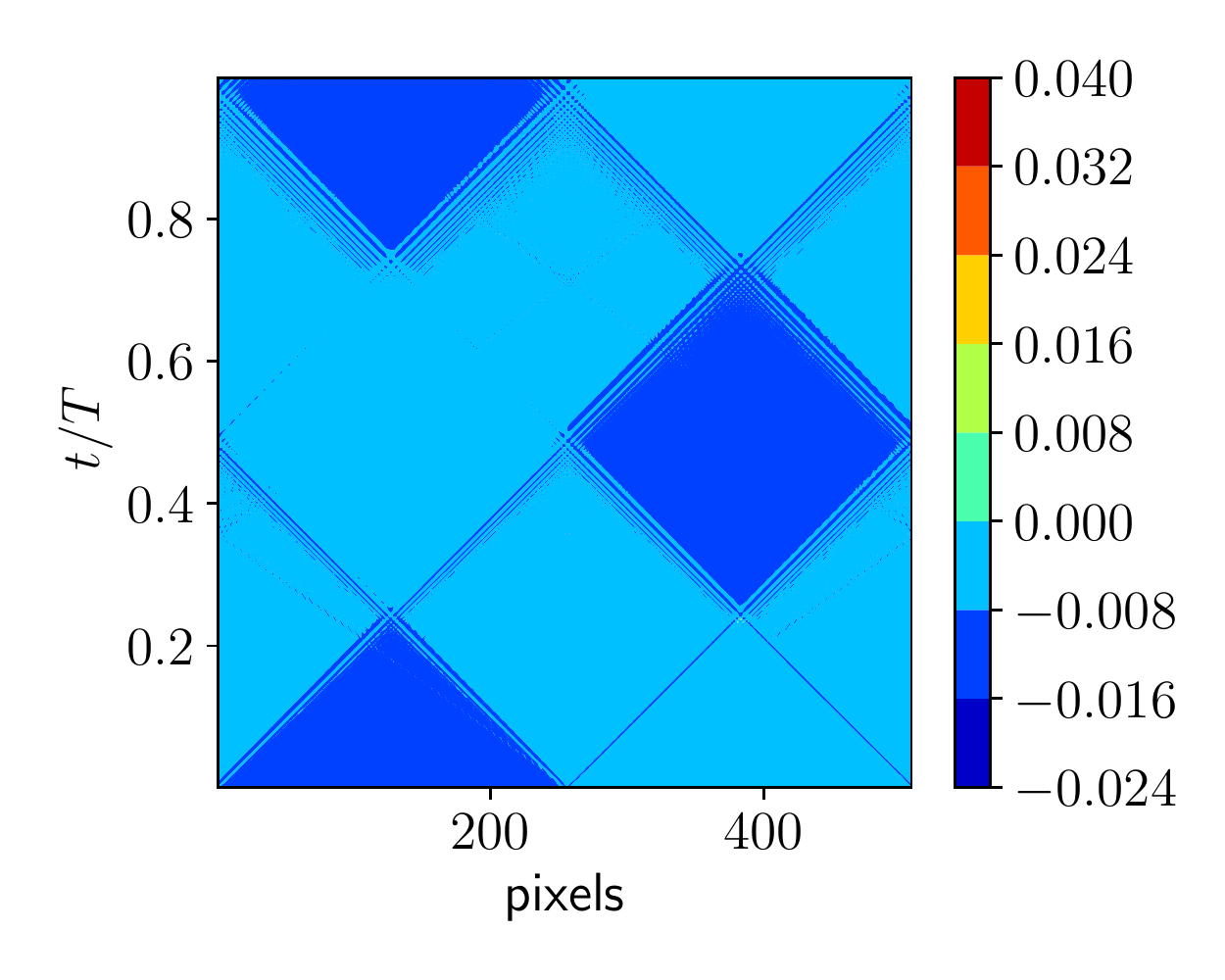}
         \caption{$f=16$}
    \end{subfigure}
\caption{Feature maps from filters 3, 9 and 16 of convolutional encoder's first convolutional layer}
\label{fig:CANfmapeconv1ls3_9_16}
\end{figure}

Unlike projection-based approaches, convolutional neural networks are not restricted to orthogonal bases. Thus they can select any general basis functions from an affine subspace. To demonstrate such flexibility associated with the CAE's heuristic learning process, we explore the feature maps obtained from its various layer filters. The evolution of the feature maps of filters 3, 9 and 16 of the first convolutional layer of the convolutional encoder are shown in Figs. \ref{fig:CANfmapeconv1ls3_9_16} (a), (b) and (c), respectively. These feature maps are a representation of the spatial domain but with a progressively reduced discretization along the spatial coordinate as we proceed deeper into the network. Thus, the evolution of these feature maps can be considered a representation of the actual $x-t$ plane. These figures clearly indicate that the CAE kernels can directly learn the features resembling wave characteristics. 

\begin{figure}[ht]
\centering
    \begin{subfigure}[b]{.48\textwidth}
        \centering
        \includegraphics[width=\textwidth]{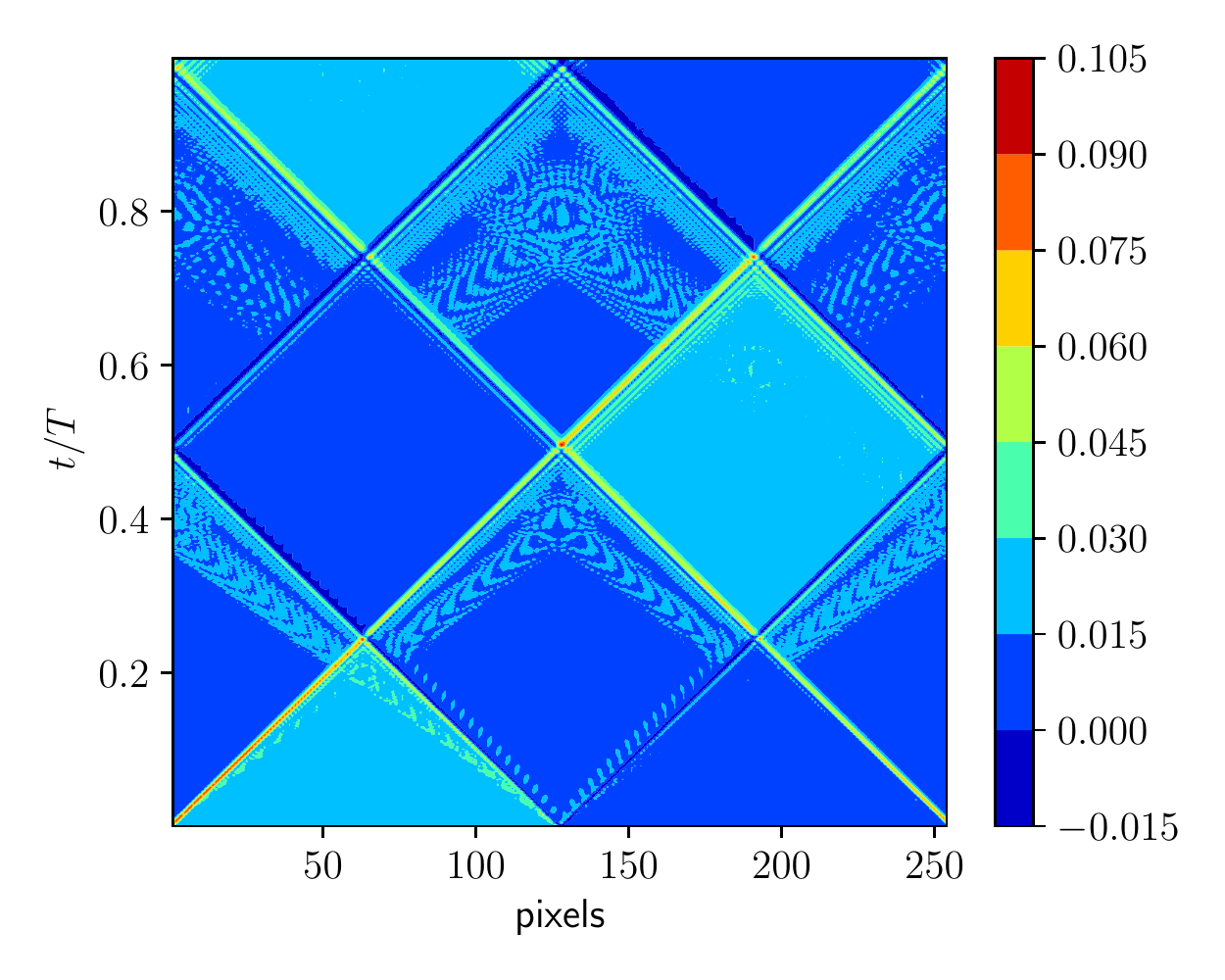}
        \caption{$f=3$}
    \end{subfigure}
    \hfill
    \begin{subfigure}[b]{.48\textwidth}
    \centering
    \includegraphics[width=\textwidth]{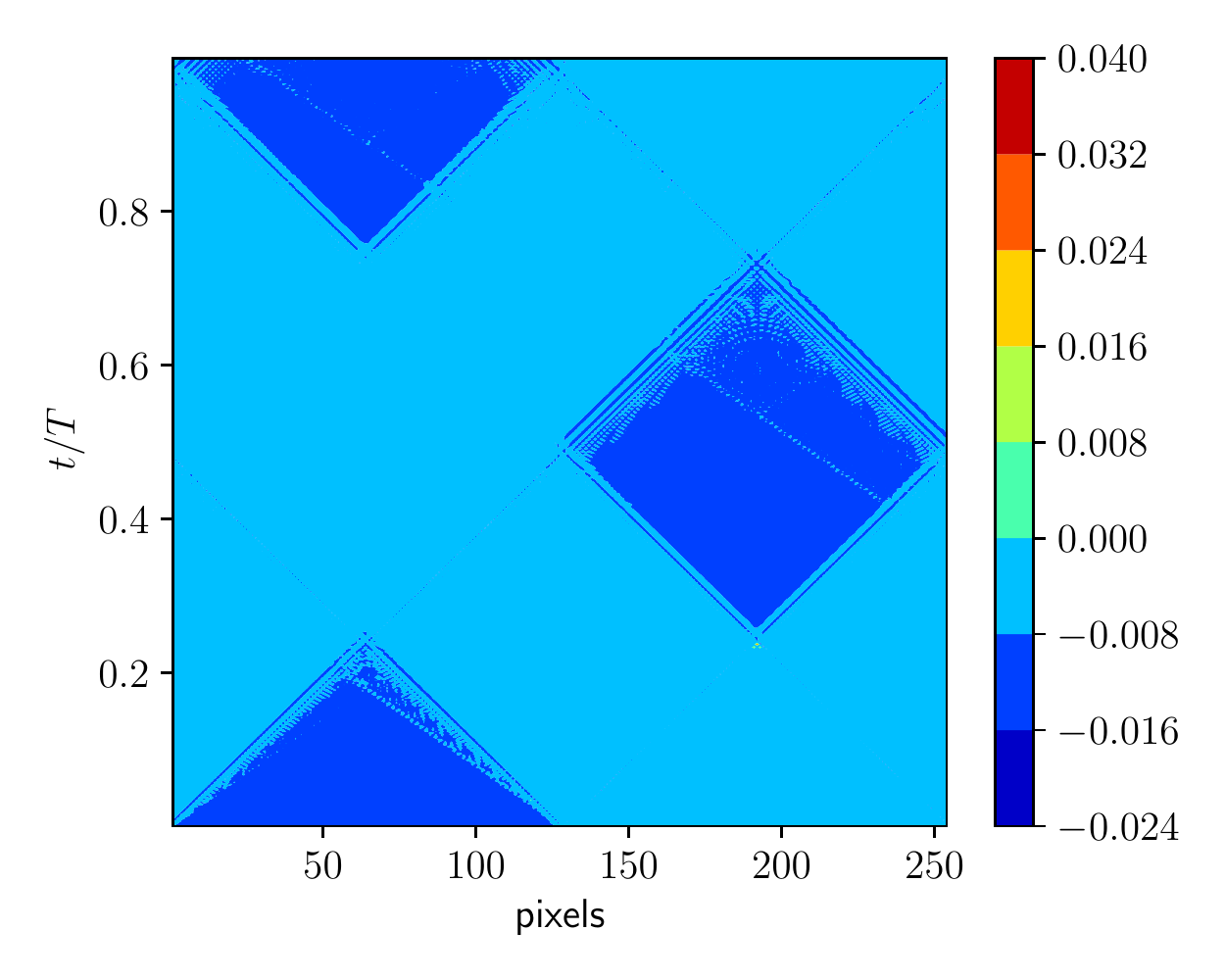}
         \caption{$f=16$}
    \end{subfigure}
\caption{Feature maps from filters 3 and 16 of convolutional encoder's first max pooling layer}
\label{fig:CANfmapemp1ls3_16}
\end{figure}

\begin{figure}[ht]
\centering
    \begin{subfigure}[b]{.48\textwidth}
        \centering
        \includegraphics[width=\textwidth]{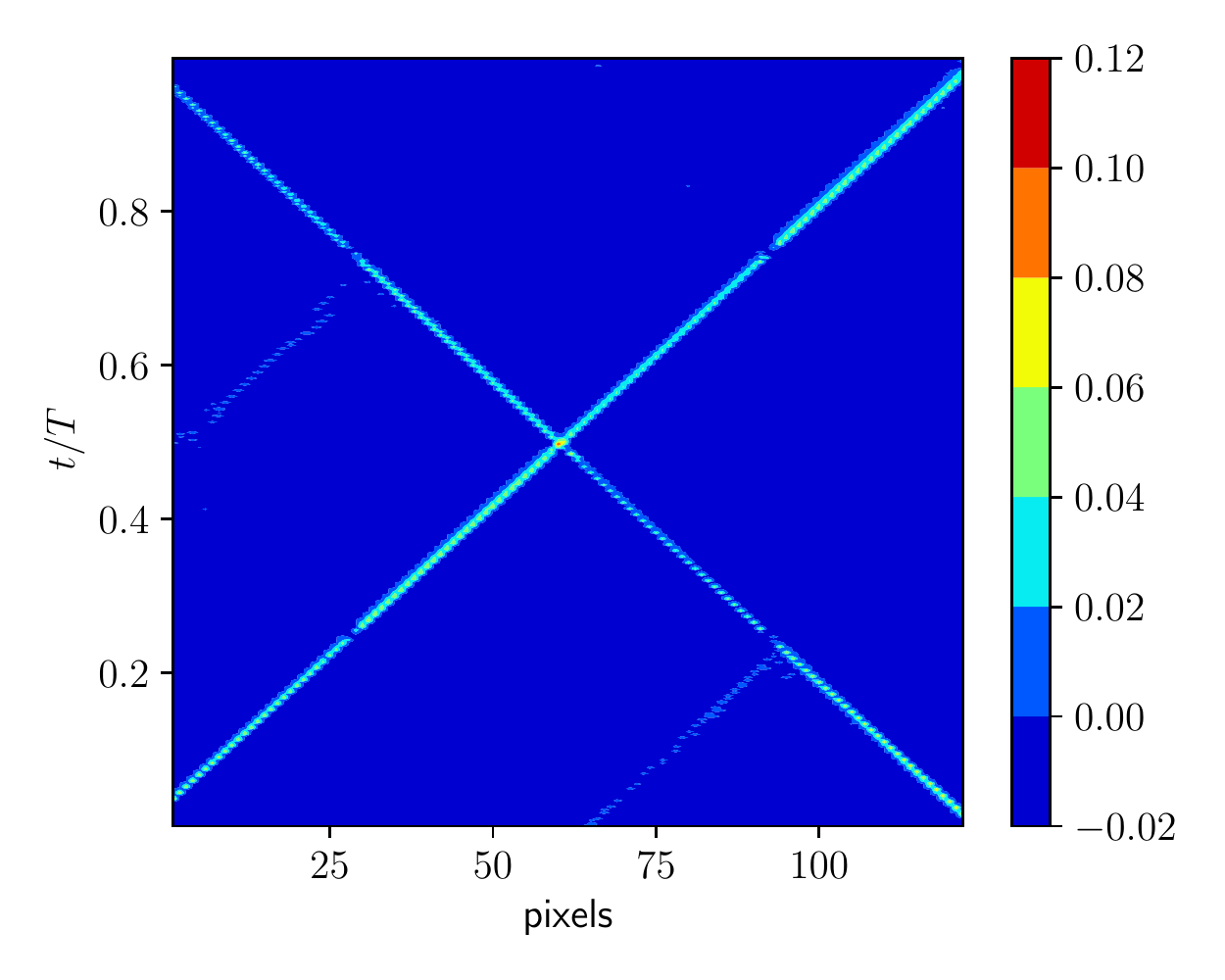}
        \caption{$f=5$}
    \end{subfigure}
    \hfill
    \begin{subfigure}[b]{.48\textwidth}
    \centering
    \includegraphics[width=\textwidth]{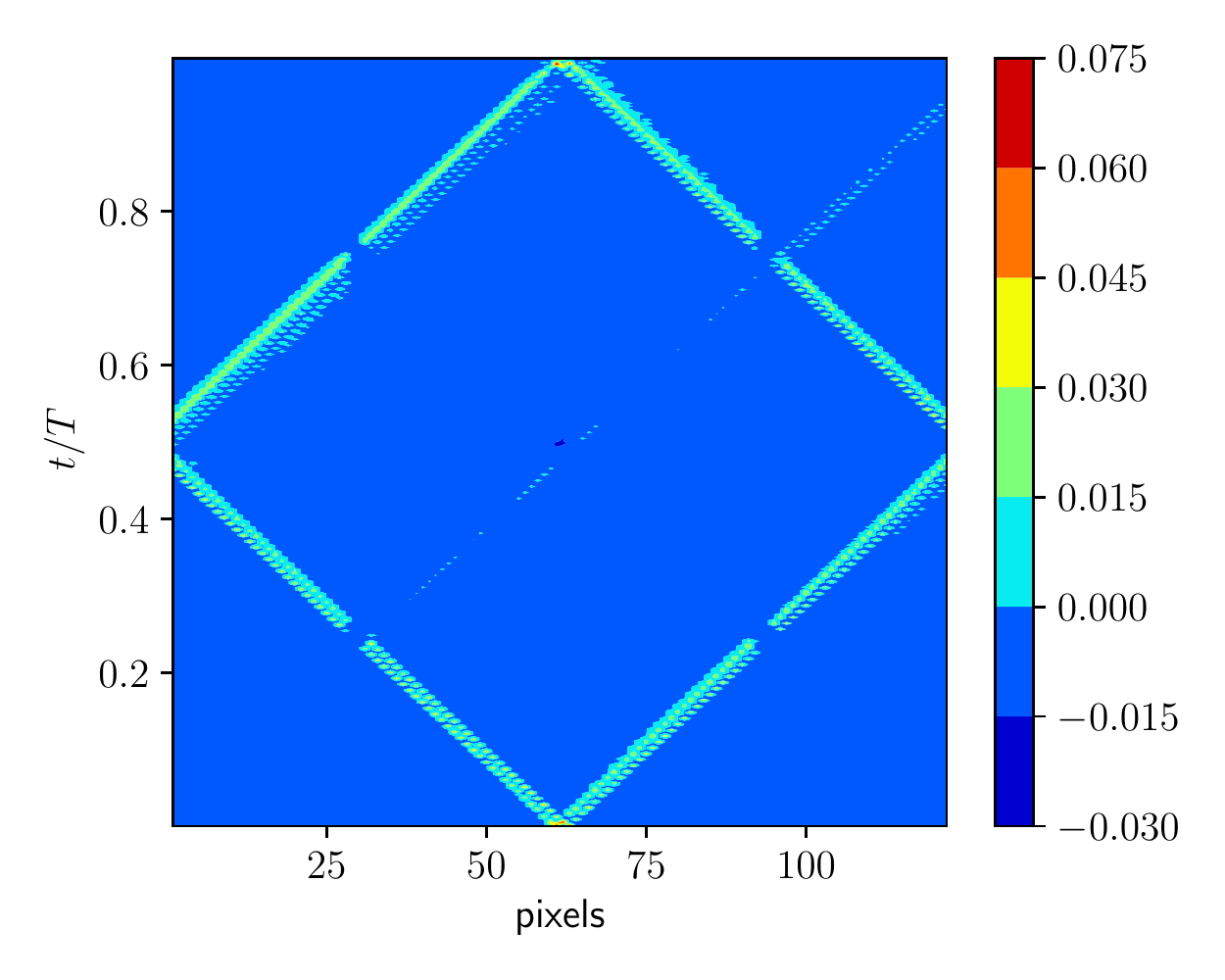}
         \caption{$f=30$}
    \end{subfigure}
\caption{Feature maps from filters 5 and 30 of convolutional encoder's third convolutional layer}
\label{fig:CANfmapeconv3ls5_30}
\end{figure}

We further observe from the feature maps of the first convolutional layer of the convolutional encoder that of filter 9 (see Fig. \ref{fig:CANfmapeconv1ls3_9_16}) directly learn the wave characteristics of the full domain wave propagation, while filters 3 and 16 focus on specific characteristic features at different location of the $x$-$t$ plane. Further refinement of such features is obtained via the scale separation of the first max pooling layer (see Fig. \ref{fig:CANfmapemp1ls3_16}). As a result, we see greater localization in the feature maps of filters 3 and 16 of this layer compared to their counterparts from the first convolution layer. Finally, after more convolution and pooling operations, filters 5 and 30 of the third convolutional layer feature maps focus on highly localized characteristic features of the wave as illustrated in Fig. \ref{fig:CANfmapeconv3ls5_30}. It is also vital to notice that the weights associated with various feature spaces vary significantly in magnitude indicating the relative participation of these features in the wave propagation dynamics. Thus, the hierarchical network of the convolutional and pooling layers leads to significant scale separation via progressive spatial reduction. Such selective scale separation, a nonlinear composition of these selective features and direct operation on subspaces resembling the wave characteristics, enables the convolutional encoder to reduce high-dimensional wave propagation observables efficiently.

It was interesting to observe earlier in Fig. \ref{fig:comp_CANls32varyingnf} that the efficiency of learning a low-dimensional map also increases with the number of available filters in the various convolutional and max pooling layers of the CAE. This observation can be explained by the fact that with more available filters the encoder has greater flexibility in separating the scales of the physical wave propagation data. Thus, more selective scale separation is possible and the convolutional encoder can obtain a more optimal set of latent states on the low-dimensional manifold. It further emphasizes how the scale separation prior provided by the pooling layers enables the CAE to reduce the high-dimensional physical observables to a low-dimensional manifold. 

\subsection{Selection of efficient autoregressive evolution model}
To compare the various LSTM networks at our disposal, we train each of them separately on the baseline finite element solution. The LSTM training for this specific purpose is performed directly on the full-dimensional solution to quantify errors that arise solely during the autoregressive prediction via the LSTM. The training input was divided into 32 training batches, where each batch represents one sequence. Since the solution for each period consists of 512 snapshots, each batch/sequence thus consists of 16 time-steps. To obtain the training output, finite element method results were obtained for one period but from time $t=T/32$  to $t=T + T/32$, where $T$ is the time period. The training output also consisted of 32 batches over one period of data. Thus, the training output batches correspond to each input training batch but are shifted by $T/32$. The evolution from input to output batches of the period is learned by the LSTM architecture. 

\begin{figure}[ht]
\centering
    \begin{subfigure}[b]{.48\textwidth}
        \centering
        \includegraphics[width=\textwidth]{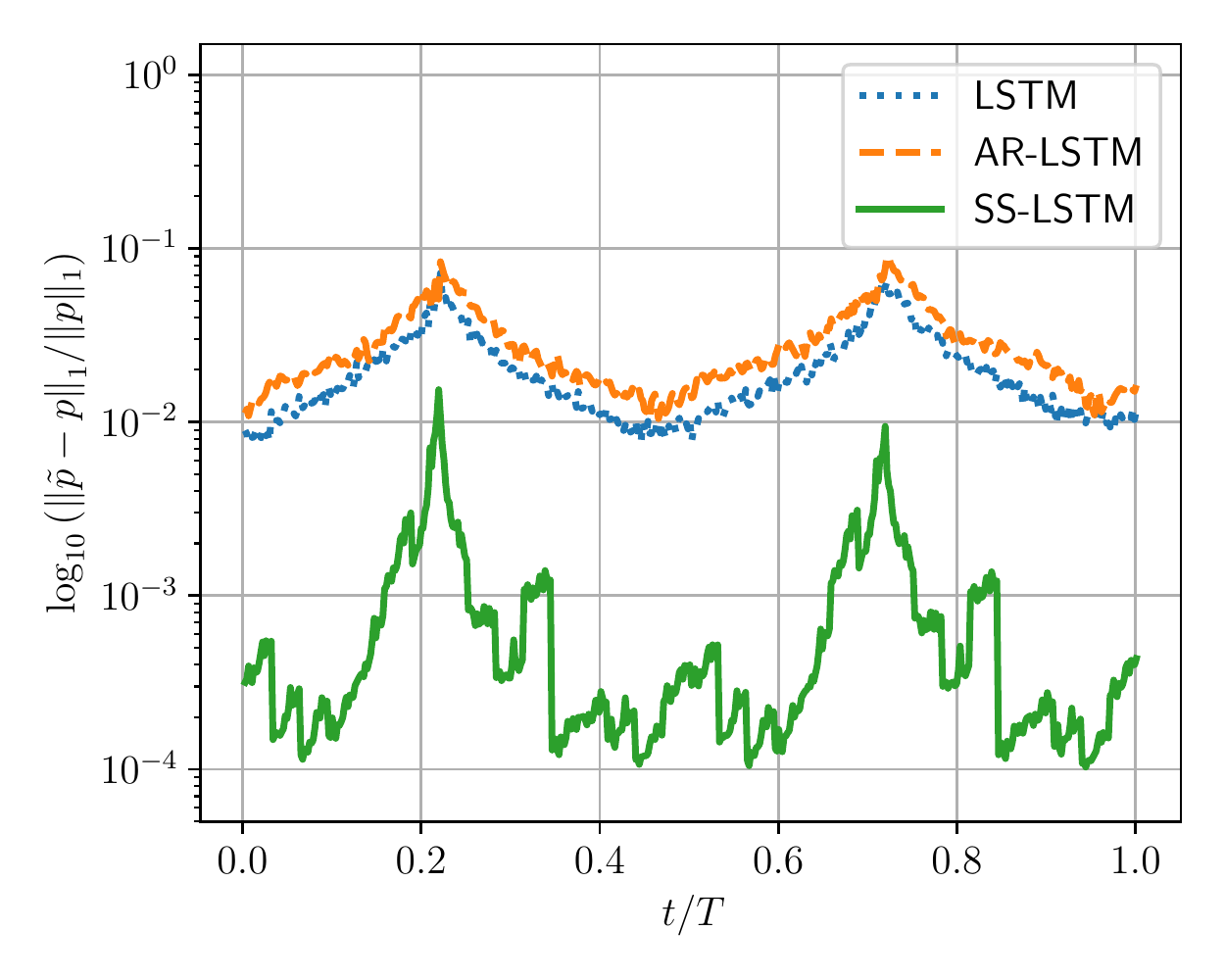}
        \caption{Training}
    \end{subfigure}
    \hfill
    \begin{subfigure}[b]{.48\textwidth}
    \centering
    \includegraphics[width=\textwidth]{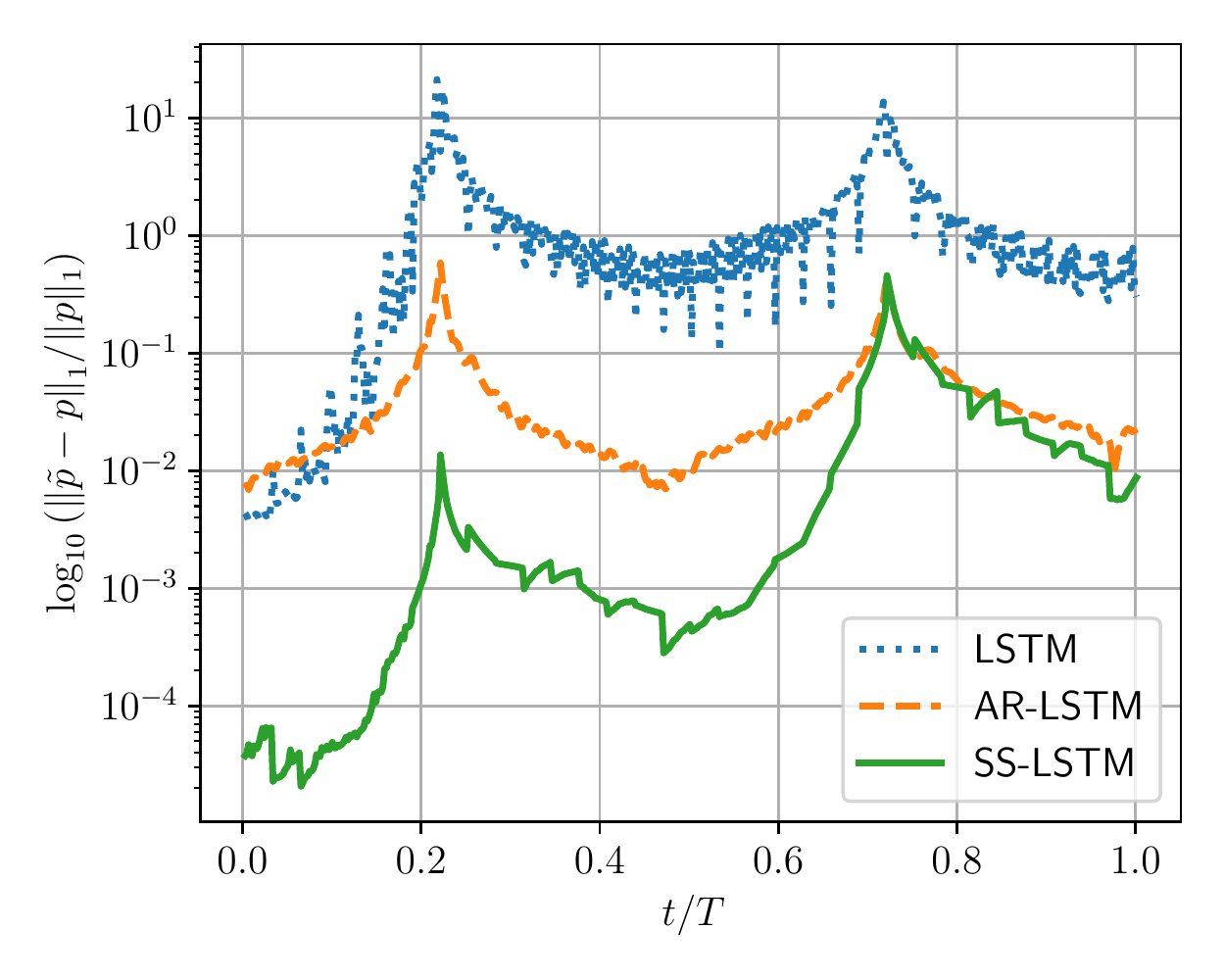}
         \caption{Testing}
    \end{subfigure}
\caption{Comparison of relative $L_1$ training and testing errors of plain LSTM, AR-LSTM and SS-LSTM networks}
\label{fig:comp_LSTM}
\end{figure}

The training results for each of the three networks are compared in Fig. \ref{fig:comp_LSTM} (a). The results show that the SS-LSTM performs the best during training with extremely small relative $L_1$ training error norms. On the other hand, both the plain and AR-LSTM trains with an order higher $L_1$ error compared to the SS-LSTM. However, the mean training errors over the whole period for these two networks remain less than 5\%. Furthermore, for all the three cases the errors peak only at $t=0.25T$ and $t=0.75T$. The peak coincides with the interference of characteristics associated with wave discontinuities (see Fig. \ref{fig:1D_discont_soln_xs0}), which are extremely difficult to predict with great accuracy. The three networks were trained with a comparable number of trainable parameters as indicated in Table \ref{tab:net_param_train_prop}. The training for all three cases is considered satisfactory, although the training time of the AR-LSTM is significantly longer than the other two networks. This finding indicates that autoregressive LSTM training takes larger computational time than plain LSTM training. However, learning the whole output sequence in a single shot in the SS-LSTM enables it to perform autoregressive modeling at a higher computational expense than plain LSTM but significantly lower than the AR-LSTM.

\begin{table}[ht]
\caption{Training time and model size for various LSTM networks}
\begin{center}
\begin{tabular}{lccc}\hline
Training parameters & LSTM & AR-LSTM & SS-LSTM \\
\hline
$N_h$ & 200 & 200 & 64\\
Trainable parameters & $6.74 \times 10^5$ & $6.74 \times 10^5$ & $6.81 \times 10^5$ \\
Training time (GPU seconds) & 380 & 3900 & 810 \\ \hline
\end{tabular}
\label{tab:net_param_train_prop}
\end{center}
\end{table}

The autoregressive prediction capability of the three trained LSTM models was subsequently tested for the next period of the data. Since the system is periodic, there is little difference between the training and test data sets. However, during the prediction phase, each network is initialized with a sequence spanning $t=T/32$. Subsequently, the remaining prediction into the horizon is obtained by iteratively feeding the predicted sequence as input for the next prediction. On the other hand, the network was fed the whole period data as a set of sequences during the training and trained to predict the next set of sequences. Thus, the iterative prediction during testing will demonstrate if errors accrue during the autoregressive prediction of the dynamics over a long time horizon. The relative $L_1$ errors in the predicted pressure fluctuations $\tilde{p}$ over the domain for the whole period, obtained with the various LSTM networks, are presented in Fig. \ref{fig:comp_LSTM} (b). We can see that errors in plain LSTM predictions begin increasing very quickly after a few iterations of its operation and generates spurious results (extremely large order of errors) after $t=0.2T$. The AR-LSTM has a much lower error which increases only during the interference of discontinuities at $t=0.25T$ and $t=0.75T$ but has a mean of 8\% over the period. The SS-LSTM initially has one order lower prediction error than the AR-LSTM but eventually reaches similar relative error values by $t=0.7T$. However, its mean error over the period is less than 5\%, which we consider the best-performing LSTM model. Subsequently, CRAN studies in this article will be performed with the SS-LSTM architecture.

The poor prediction performance of the plain LSTM compared to its SS-LSTM and AR-LSTM counterparts can be attributed to its lack of autoregressive modeling. During training, the plain LSTM network is only trained to learn the mapping between units of the input sequence to their corresponding units of their output sequence (please see Fig. \ref{fig:LSTM_architectures} (a)). 
In the absence of autoregressive modeling, this type of multi-output prediction strategy creates a disconnect between the predicted states of the plain LSTM and those states immediately preceding them. As a result of such disconnect or lack of dependency, errors quickly accrue during the iterative prediction of an evolving transport system. 

\subsection{Learning spatially distributed discontinuities via CRAN}
Herein, the results for the generalized learning capacity of the CRAN model for spatially distributed discontinuities along the domain will be discussed. To train the CRAN, temporal snapshots of the evolution of the wave propagation over a period were generated for ten different discontinuity locations in the initial condition ($x_s$ in Eq. \ref{eq:discont_IC}). Two validation sets for $x_s$ and five test sets for $x_s$ were also selected. These 17 $x_s$ values were randomly sampled from the domain and their locations were adjusted to lie on the nearest node of the finite element mesh. Thus, the same mesh could be applied to generate all 17 $x_s$ sets. The data set for each different $x_s$ had 512 temporal snapshots. The CRAN training was separated into a CAE training and an LSTM training on the reduced-dimensional latent states obtained from the trained convolutional encoder. 

First, we discuss the CAE training. A hyperparameter tuning was performed for optimal CAE training and generalization. Several network parameters like convolutional neural network filter size, number of filters, etc., were considered for optimal network tuning. However, the most important hyperparameter considered here is the number of training epochs. The accuracy of the predicted solution compared to the target was measured via Structural Similarity Index Measure (SSIM), a statistical measure used for comparing two images or two same-dimensional data sets with pixelated representation. The formal definition of the SSIM can be obtained from article \cite{wang2004image}. An SSIM of 1.0 between two images indicates that the two images are identical. SSIM decreases from 1 as the similarity between two images decreases and 0 indicates no similarity. Since the solutions here can be represented in a pixelated format on a uniform space and time grid, SSIM is considered an adequate measure for quantifying network prediction accuracy. 

\begin{figure}[ht]
\centering
    \begin{subfigure}[b]{.48\textwidth}
        \centering
        \includegraphics[width=\textwidth]{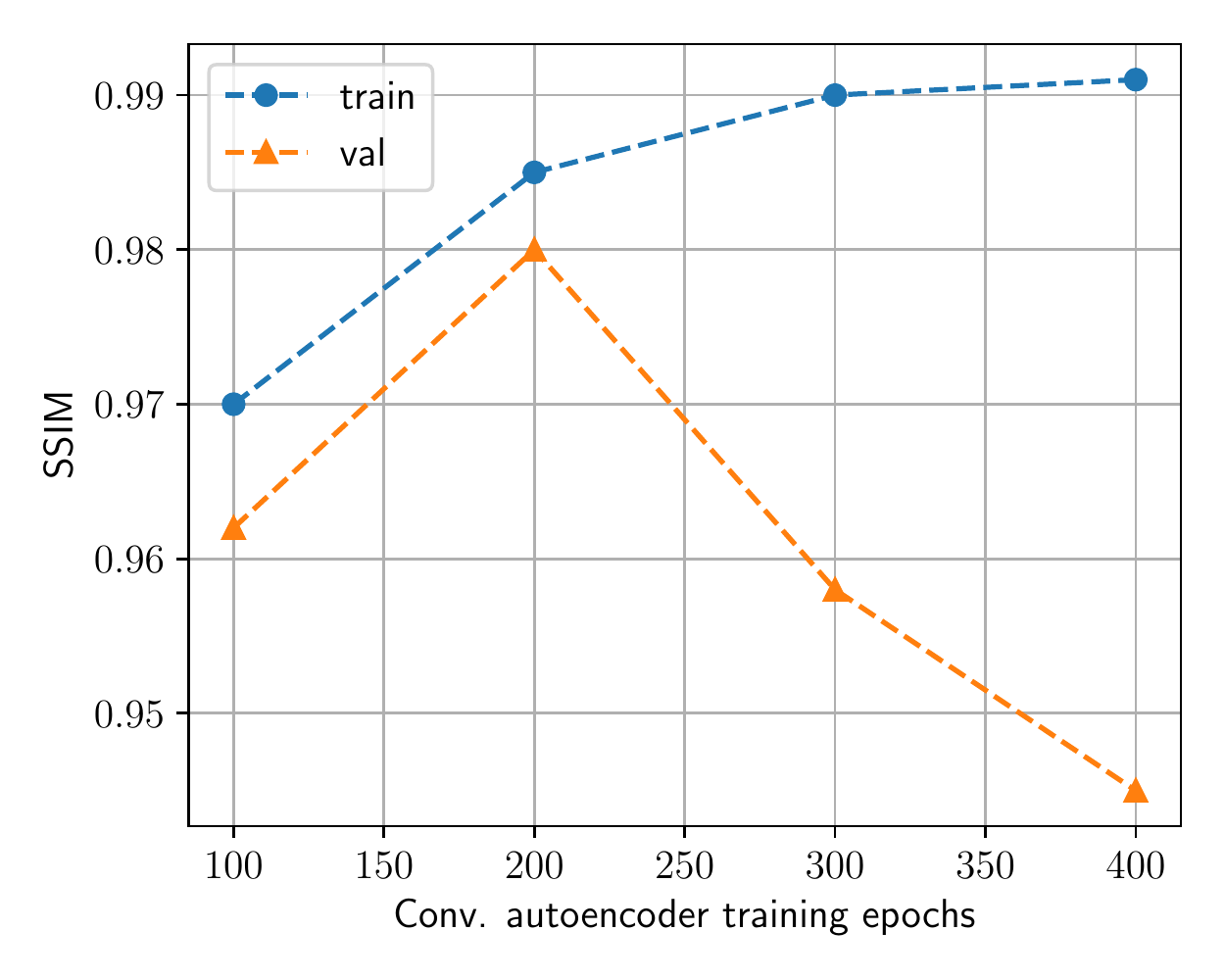}
        \caption{Convolutional autoencoder}
    \end{subfigure}
    \hfill
    \begin{subfigure}[b]{.48\textwidth}
    \centering
    \includegraphics[width=\textwidth]{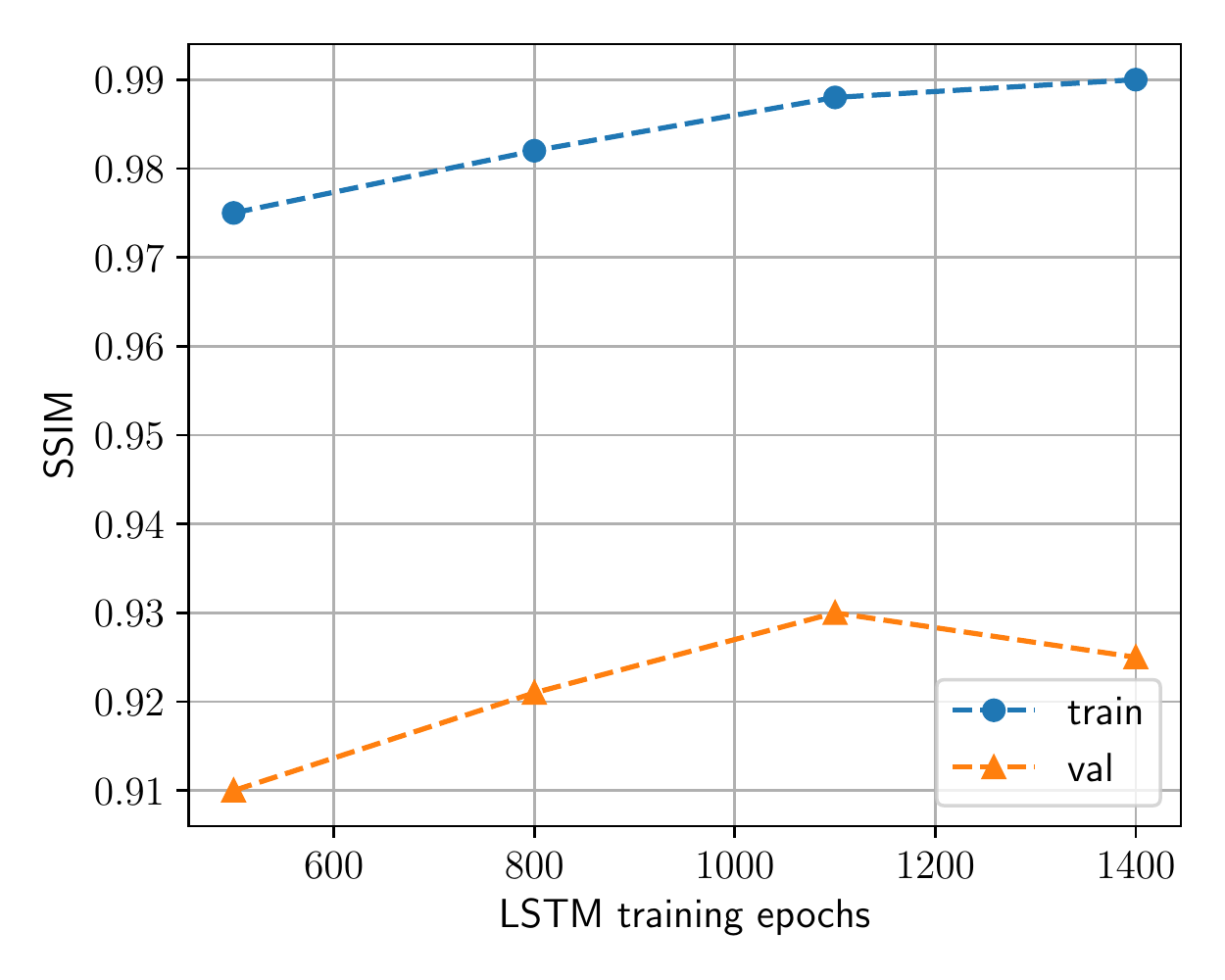}
         \caption{CRAN}
    \end{subfigure}
\caption{Effects of training epoch on network training and validation for 1D wave propagation: (a) SSIM of CAE predictions, (b) SSIM of CRAN predictions}
\label{fig:hyp_tun_epoch_1D_discont}
\end{figure}

Fig. \ref{fig:hyp_tun_epoch_1D_discont} (a) shows the change in the mean SSIM of the predicted solutions compared to the target solutions considering all the ten $x_s$ combined, with an increase in the training epochs. A similar mean SSIM is also presented for the two validation $x_s$ combined. We can observe that with 10 $x_s$ data set for training, the network shows a significant influence of overfitting. The onset of overfitting is indicated by a decrease in the validation SSIM beyond 200 training epochs. Thus, the CAE trained with 200 epochs is considered the optimally trained network.

\begin{figure}[ht]
\centering
\includegraphics[width=0.49\linewidth]{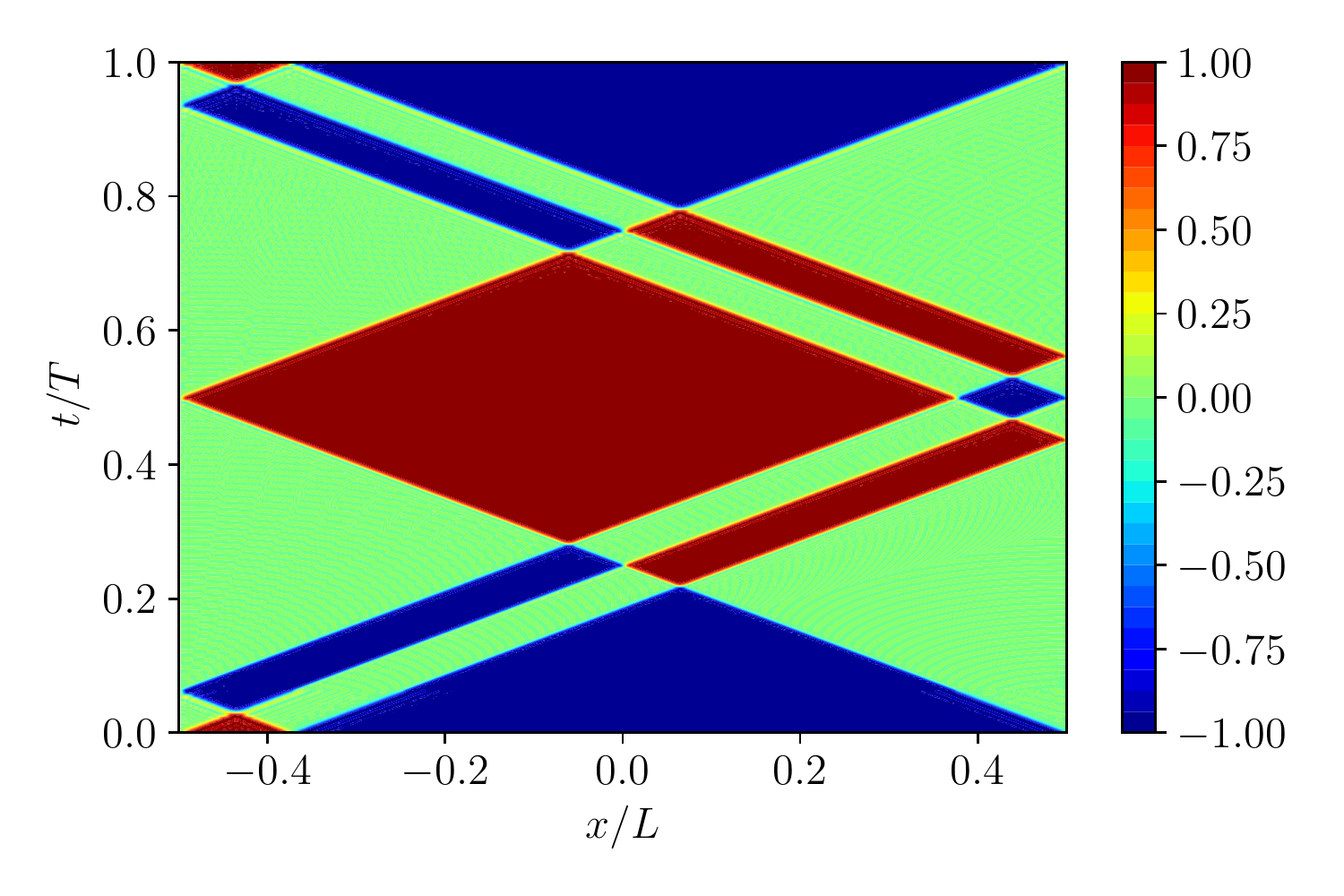}
\caption{Normalized pressure fluctuations for discontinuous initial condition obtained via finite element method: $x_s/L=-0.375$}
\label{fig:1D_discont_xsmp375}
\end{figure}

\begin{figure}[ht]
\centering
    \begin{subfigure}[b]{.32\textwidth}
        \centering
        \includegraphics[width=\textwidth]{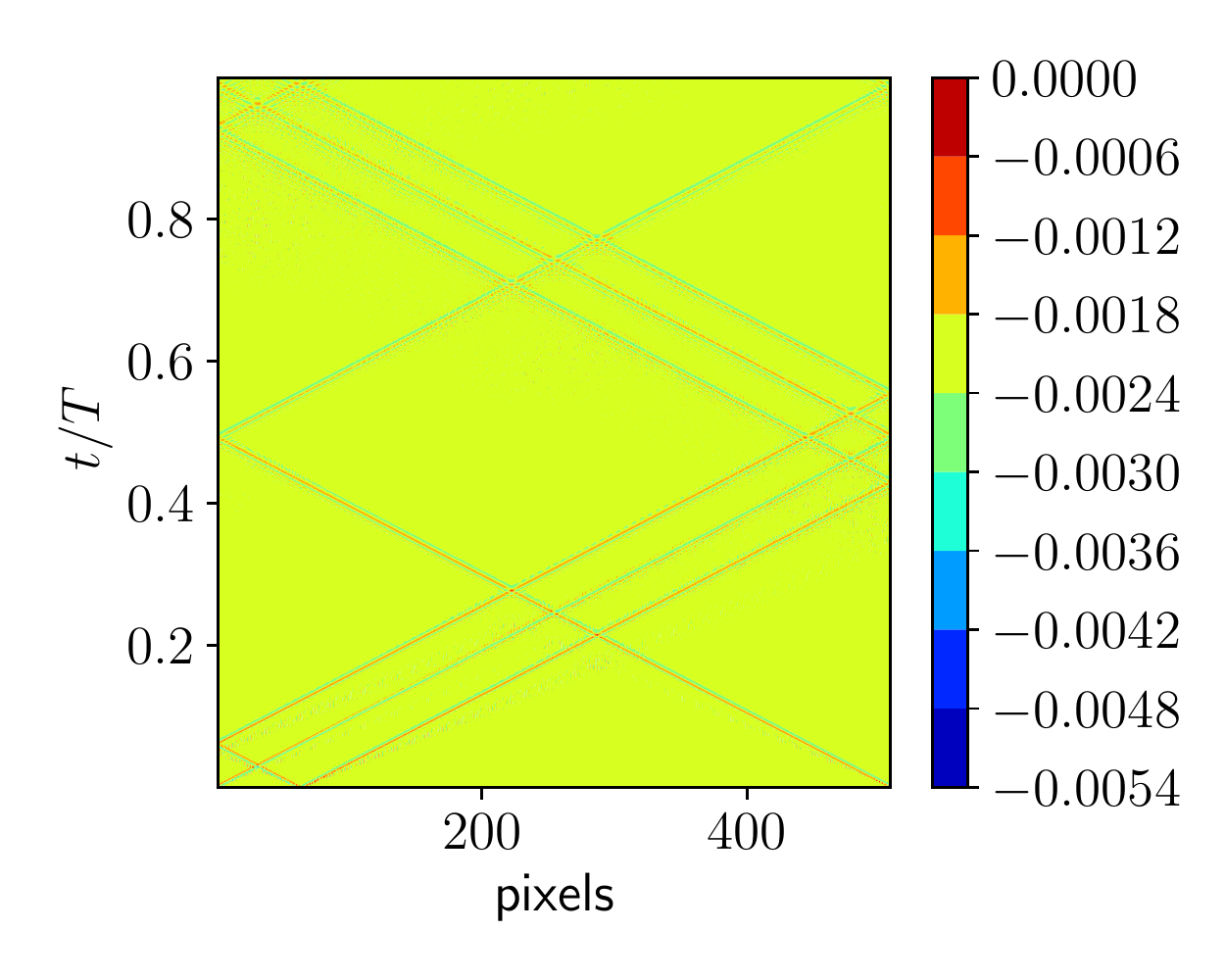}
        \caption{$f=6$}
    \end{subfigure}
    \hfill
    \begin{subfigure}[b]{.32\textwidth}
    \centering
    \includegraphics[width=\textwidth]{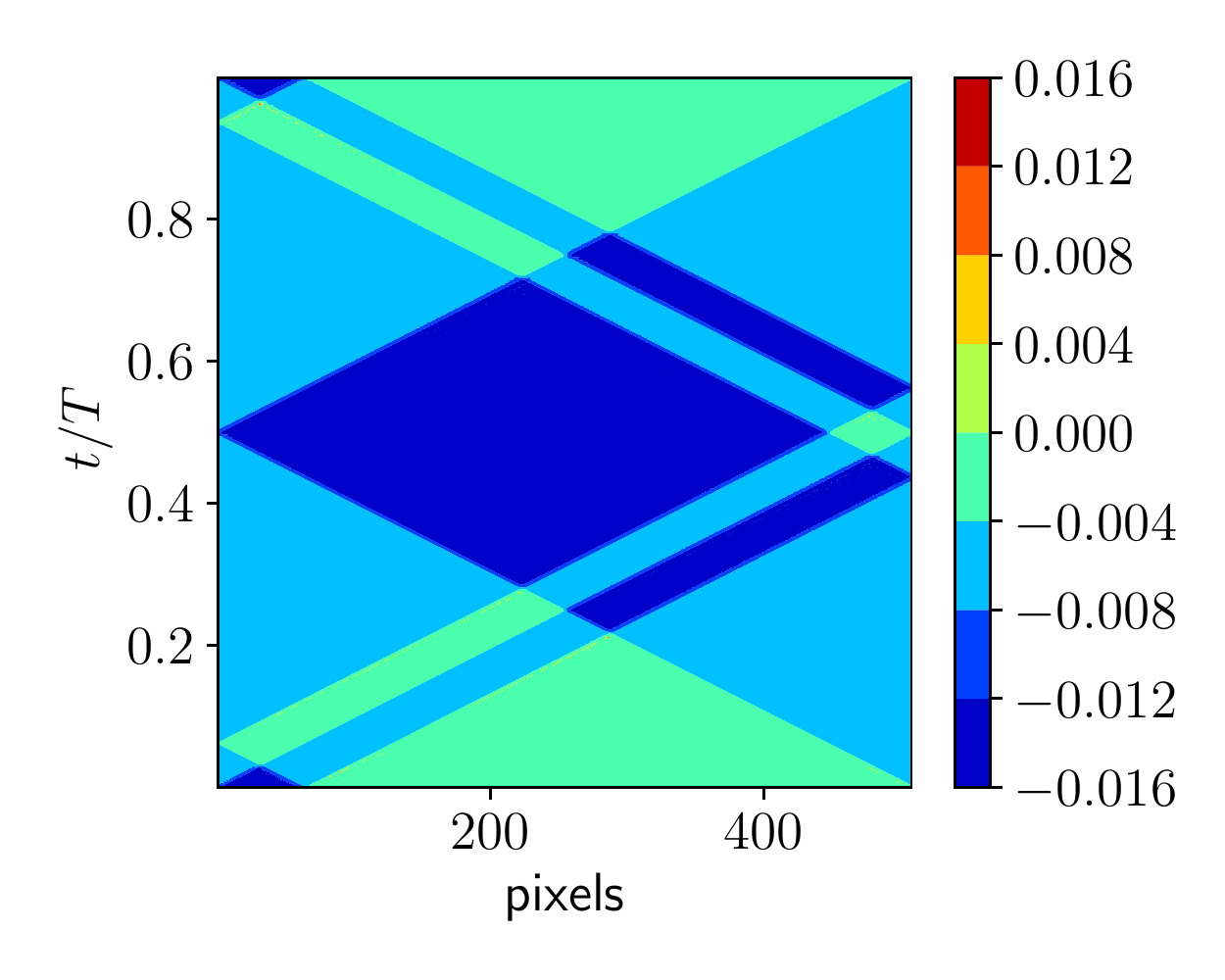}
         \caption{$f=8$}
    \end{subfigure}
    \begin{subfigure}[b]{.32\textwidth}
    \centering
    \includegraphics[width=\textwidth]{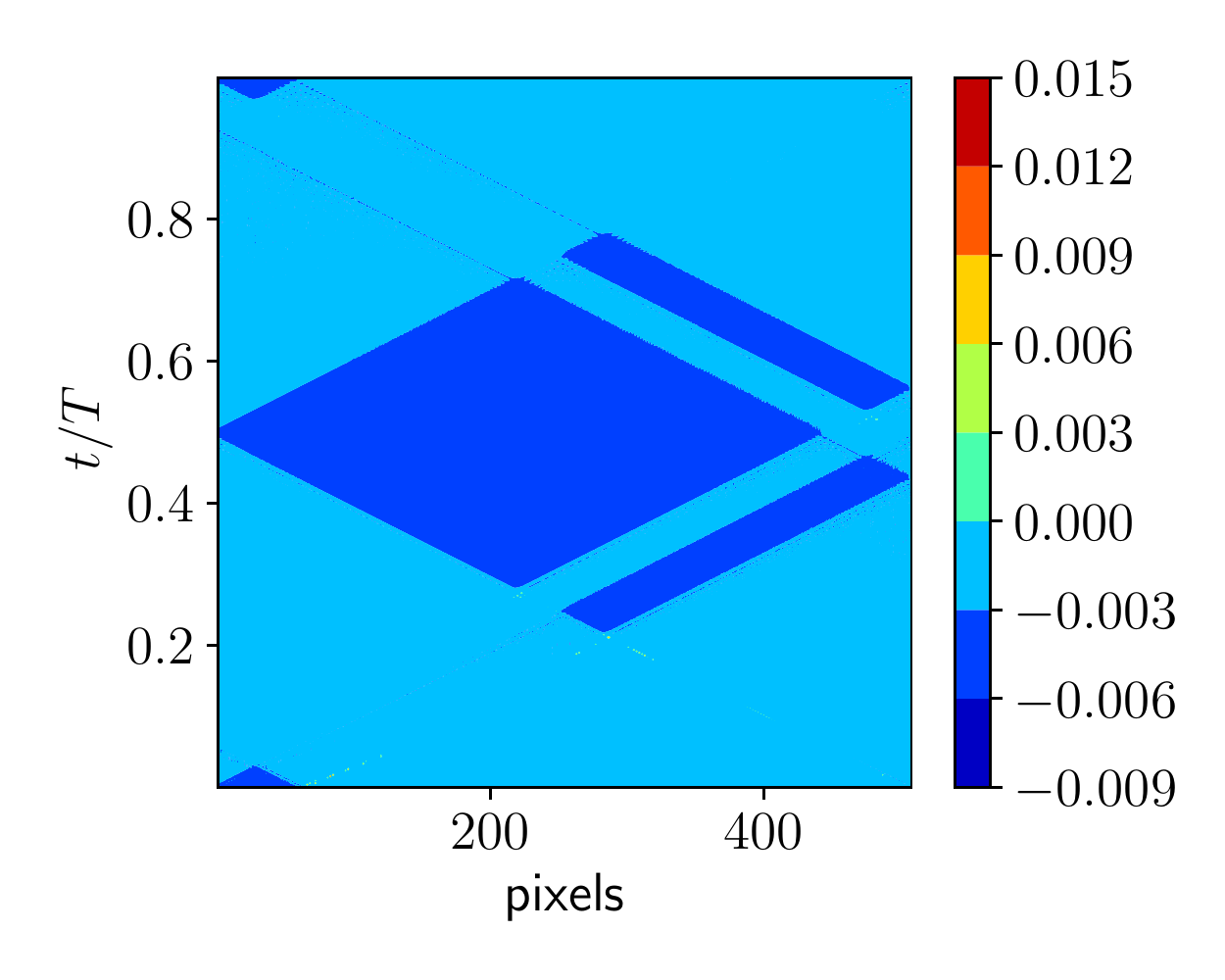}
         \caption{$f=16$}
    \end{subfigure}
\caption{Feature maps from filters 6, 8 and 16 of first convolutional layer: convolutional encoder trained with spatially varying initial conditions}
\label{fig:CAN10infmapeconv1ls6_8_16}
\end{figure}

\begin{figure}[ht]
\centering
    \begin{subfigure}[b]{.48\textwidth}
        \centering
        \includegraphics[width=\textwidth]{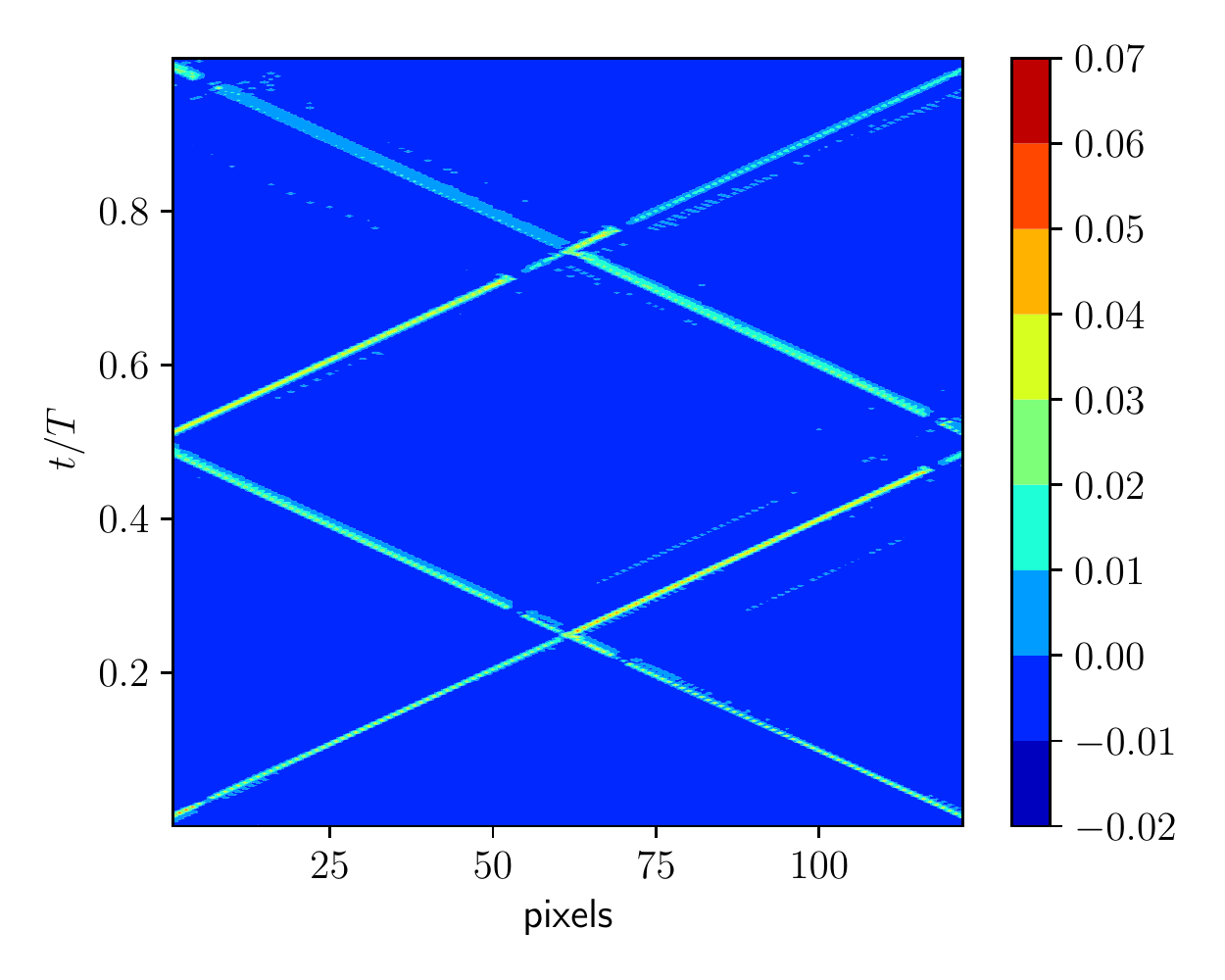}
        \caption{$f=9$}
    \end{subfigure}
    \hfill
    \begin{subfigure}[b]{.48\textwidth}
    \centering
    \includegraphics[width=\textwidth]{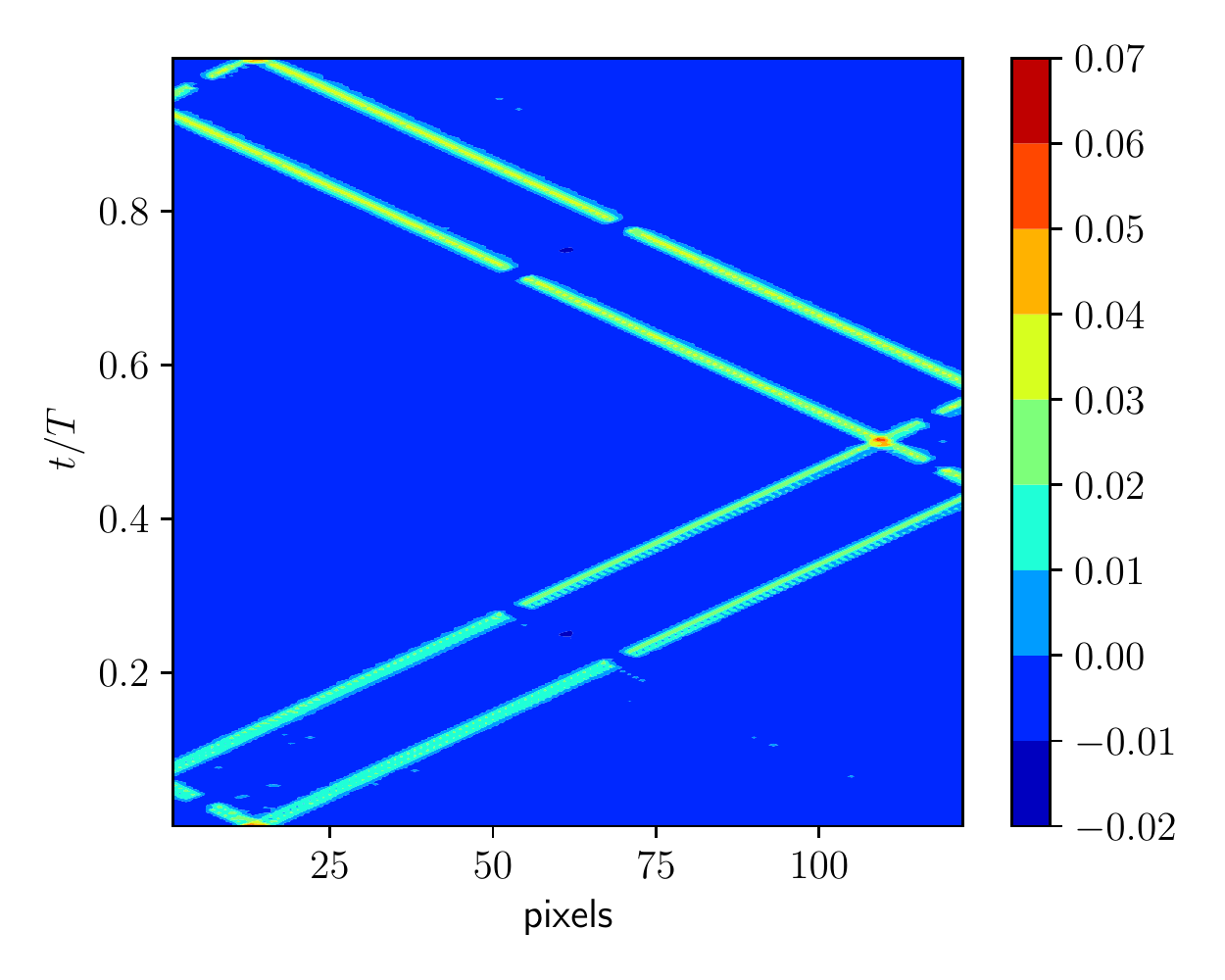}
         \caption{$f=39$}
    \end{subfigure}
\caption{Feature maps from filters 9 and 39 of third convolutional layer: convolutional encoder trained with spatially varying initial conditions}
\label{fig:CAN10infmapeconv3ls9_39}
\end{figure}

We will now physically interpret the generalized prediction capability of the early-stopped convolution autoencoder by exploring its feature maps for one of the validation sets. We specifically consider the validation set of $x_s/L=0.375$, the true solution being shown in Fig. \ref{fig:1D_discont_xsmp375}. First, we explore the feature maps of filters 6, 8 and 16 of the first convolutional layer of the convolutional encoder in Figs. \ref{fig:CAN10infmapeconv1ls6_8_16} (a), (b) and (c), respectively. It can be observed that the feature map of filter 8 represents the full domain characteristics and filters 6 and 16 focus on localized characteristic features of the wave propagation in the domain. The feature maps of the third convolutional layer of the convolutional encoder are also presented in Figs. \ref{fig:CAN10infmapeconv3ls9_39} (a) and (b), for filters 9 and 39, respectively. It is observed that they focus on highly localized wave characteristics features. It can be easily deduced that the multi-resolution characteristics features observed in the various feature maps of the first and third convolutional encoder layer belong to the wave propagation phenomena observed in Fig. \ref{fig:1D_discont_xsmp375}. This indicates that the network is able to predict generalized wave propagation physics initiated by spatially varying initial conditions and not merely memorize the training cases. 

Next, we discuss the SS-LSTM training. Since we are going to train the SS-LSTM on the reduced-dimensional latent states obtained from the trained convolutional encoder, there is no specific target latent data sets. Thus, we instead inspect the accuracy of the LSTM training and validation sets on the reconstructed CRAN predictions and the target, for both training and validation data sets. Since we are keeping the CAE network fixed, any change in the errors will solely be due to the LSTM network. To operate the CRAN, the 520 snapshots of the target solution is divided into 32 batches/sequences with each sequences containing 16 snapshots and spanning $t=T/32$. The SS-LSTM parameters were trained by minimizing the mean square error between the reconstructed CRAN predictions and the target over all the batches and all spatial location of initial conditions. During validation and subsequent testing, the CRAN was operated in the prediction phase. In the prediction phase the CRAN was initiated with first batch/sequence of the target spanning $t=T/32$ and autoregressive prediction was obtained for a complete period from $t=T/32$ to $t=T+T/32$. The SSIM of the predicted CRAN solutions compared to the target solutions during training considering all the ten $x_s$ combined is shown in Fig. \ref{fig:hyp_tun_epoch_1D_discont} (b), with an increase in the training epochs. A similar mean SSIM is also presented for the two validation $x_s$ combined. Similar to the CAE training, we also observe overfitting during the LSTM training. The validation set indicates that overfitting sets in after 1100 training epochs. Thus, the SS-LSTM network trained for 1100 epochs is considered optimal. The parameters of the trained convolutional encoder, the LSTM network and the convolutional decoder are shown in Tables \ref{tab:CRAN_enc_1dwave}, \ref{tab:CRAN_prop_1dwave} and \ref{tab:CRAN_dec_1dwave}, respectively, for a sequence containing 16 snapshots. 

\begin{figure}[ht]
\centering
\includegraphics[width=0.49\linewidth]{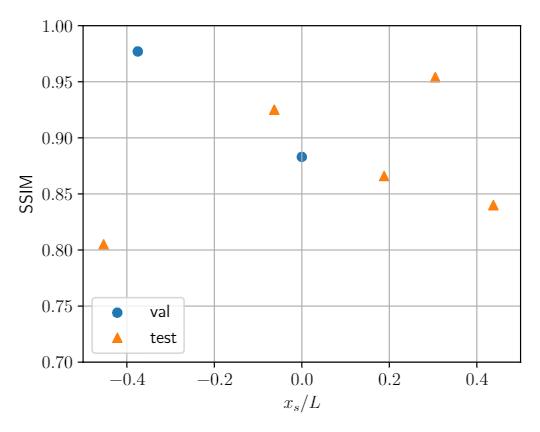}
\caption{Structural similarity index measure of CRAN's wave propagation prediction for all validation and test sets}
\label{fig:SSIM_val_test}
\end{figure}

The CRAN network with the trained CAE and SS-LSTM was next employed on the test $x_s$ locations to predict the evolution of the wave propagation over a whole period. The SSIMs for the five test $x_s$ sets and the two validation sets are presented in Fig. \ref{fig:SSIM_val_test}. We can observe that except for the case of $x_s=-0.47L$, the predicted solutions show 85\% or more similarity to the target solutions. 

\begin{figure}[ht]
\centering
    \begin{subfigure}[b]{.32\textwidth}
        \centering
        \includegraphics[width=\textwidth]{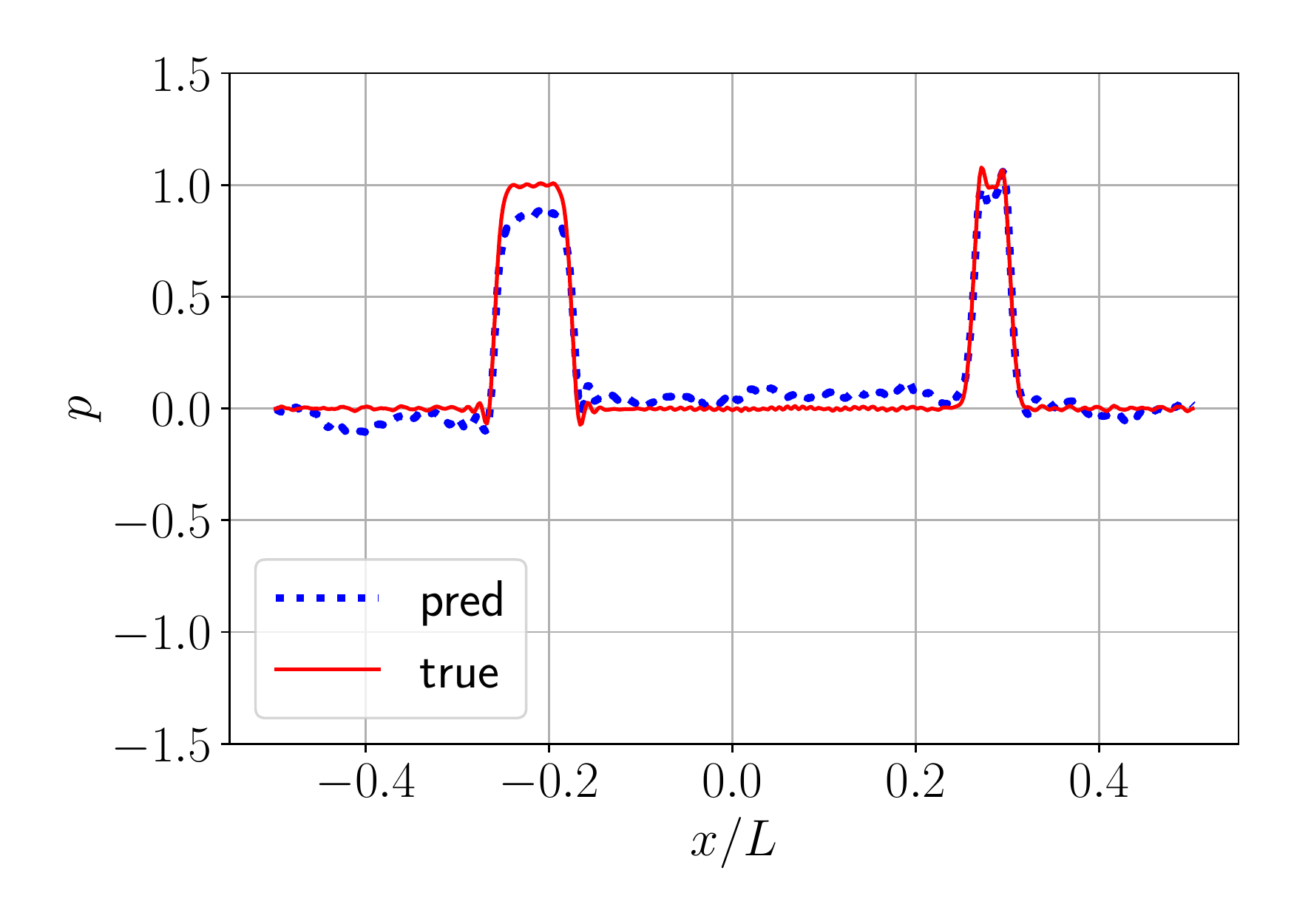}
        \caption{$t=0.38T$}
    \end{subfigure}
    \hfill
    \begin{subfigure}[b]{.32\textwidth}
    \centering
    \includegraphics[width=\textwidth]{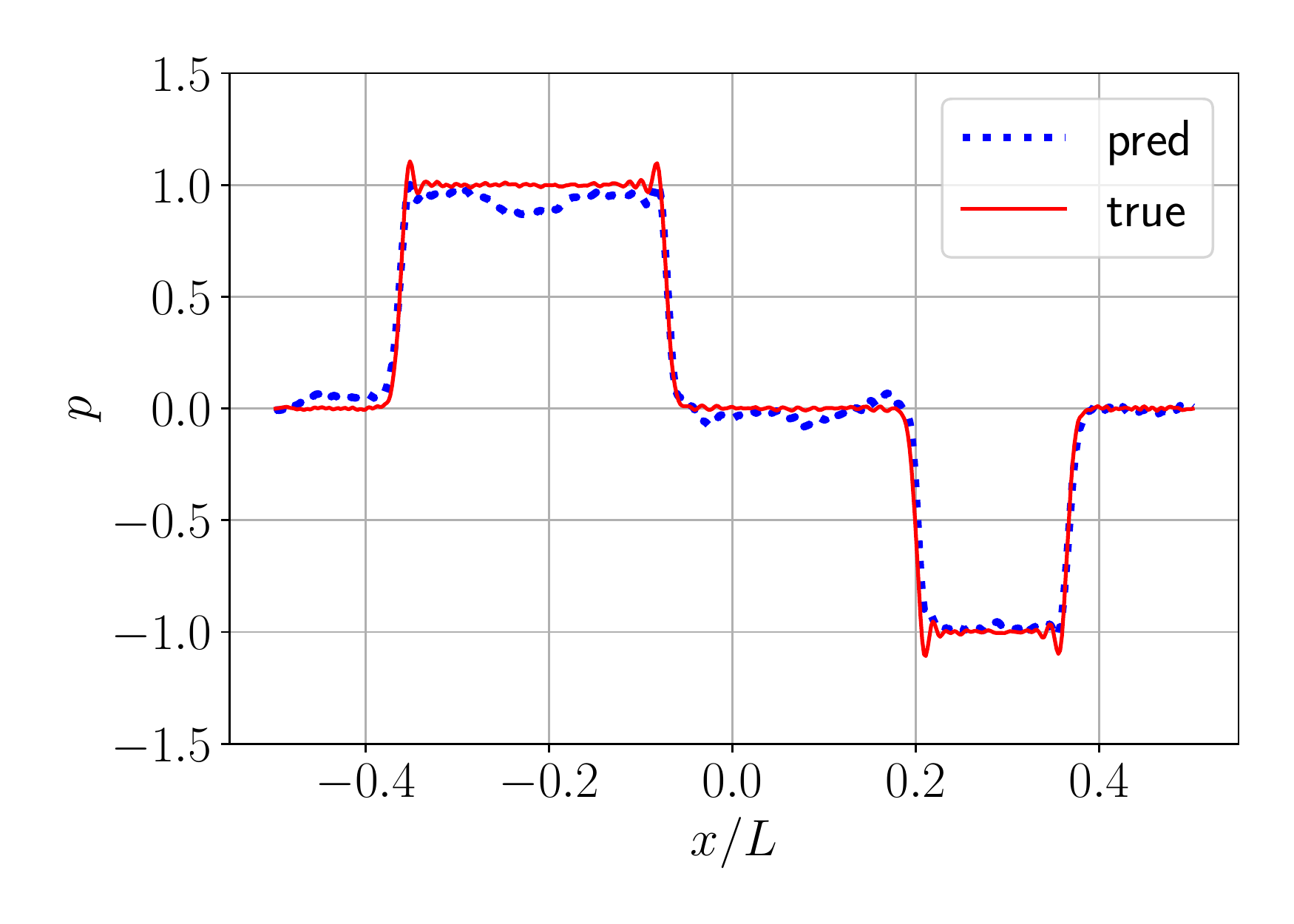}
         \caption{$t=0.57T$}
    \end{subfigure}
    \hfill
    \begin{subfigure}[b]{.32\textwidth}
        \centering
        \includegraphics[width=\textwidth]{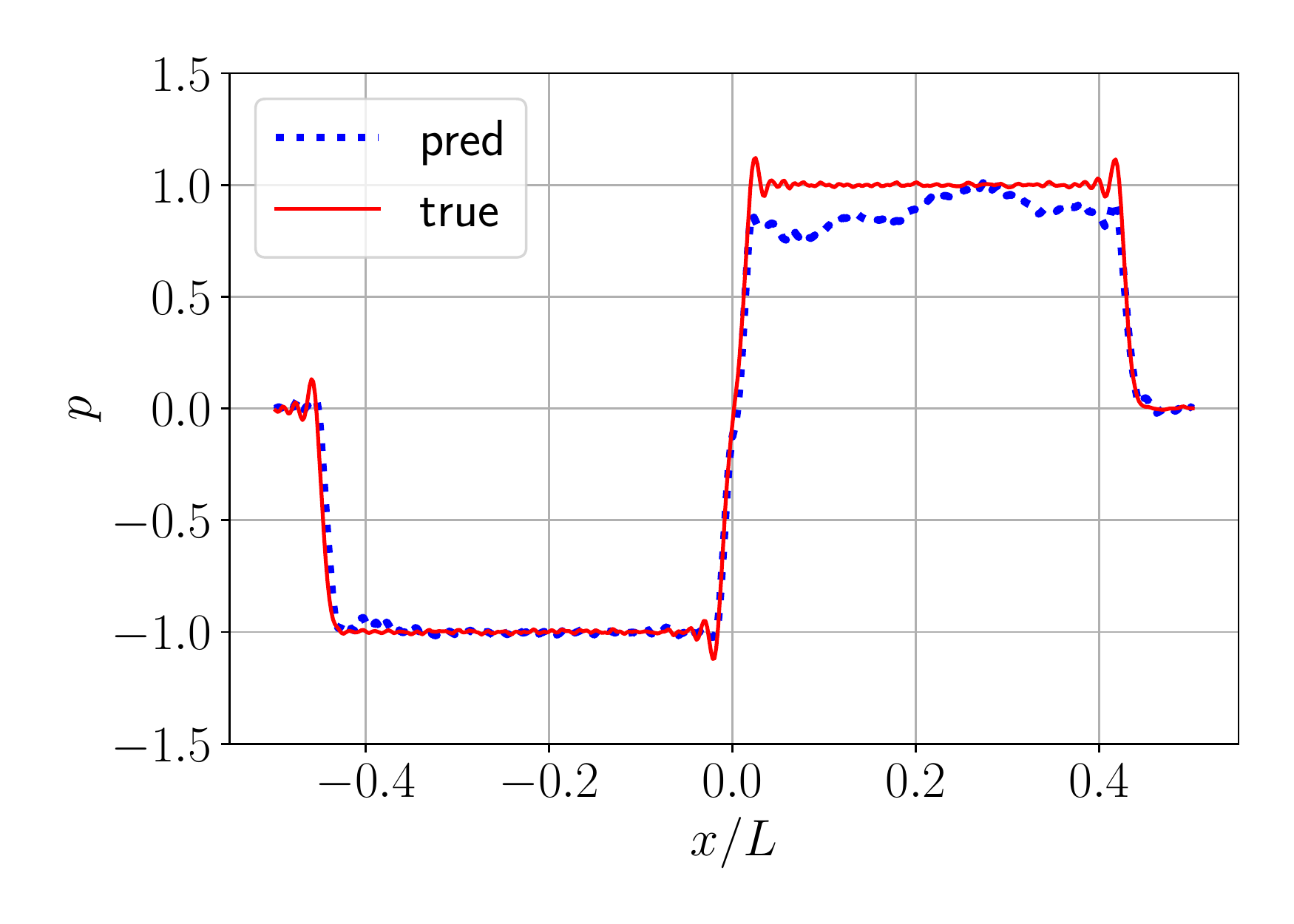}
        \caption{$t=0.76T$}
    \end{subfigure}
\caption{True and CRAN predictions of 1D wave propagation at various time instants with $x_s/L = -0.0625$: (a) $t=0.38T$, (b) $t=0.57T$, (c) $t=0.76T$}
\label{fig:comp_1D_discont_xsmp0625_vartbyT}
\end{figure}

\begin{figure}[ht]
\centering
    \begin{subfigure}[b]{.48\textwidth}
        \centering
        \includegraphics[width=\textwidth]{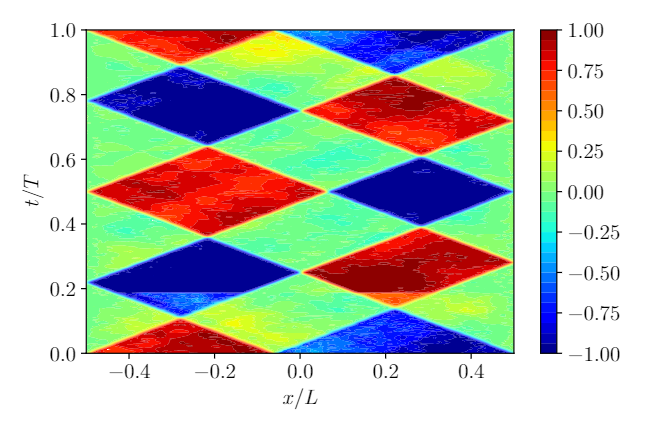}
        \caption{CRAN prediction}
    \end{subfigure}
    \hfill
    \begin{subfigure}[b]{.48\textwidth}
    \centering
    \includegraphics[width=\textwidth]{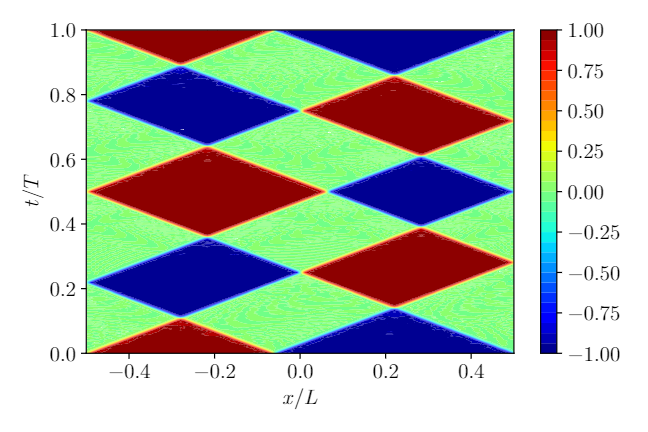}
         \caption{True solution}
    \end{subfigure}
\caption{Comparison of 1D wave propagation predictions from CRAN with true solutions for: $x_s/L = 0.0625$}
\label{fig:comp_1D_discont_xsmp0625}
\end{figure}

To further investigate the nature of the predicted results we compare the target (`true') and predicted (`pred') solutions at several temporal phases of the evolution for $x_s=-0.0625L$, in Fig. \ref{fig:comp_1D_discont_xsmp0625_vartbyT}. The results show some differences between the peak target's magnitude and predicted solutions for $t=0.38T$ and $t=0.76T$. However, the relative $L_1$ error between of the prediction is 13\% between $x/L=-0.26$ to $x/L=-0.18$ for $t=0.38T$. For $t=0.76T$, the peak prediction $L_1$ error reaches 20\% but only between $x/L=0.05$ to $x/L=0.18$. For all other peak locations for $t=0.38T, 0.57T$ and $0.76T$, the $L_1$ error lies within 10\%. More importantly, the discontinuity propagation in the predicted solution closely matches in both structure and wave speed with the target solution. This can be observed more clearly on projecting the results on the $x-t$ plane, as shown in Fig. \ref{fig:comp_1D_discont_xsmp0625}. We can see the differences in the magnitude of the target and predicted solutions at some time steps but the predicted wave propagation always follows the characteristic lines. Such characteristics are an integral part of the solutions of any hyperbolic partial differential equation, including wave propagation. Thus, the predicted solutions show physical consistency.  

\begin{figure}[ht]
\centering
    \begin{subfigure}[b]{.48\textwidth}
        \centering
        \includegraphics[width=\textwidth]{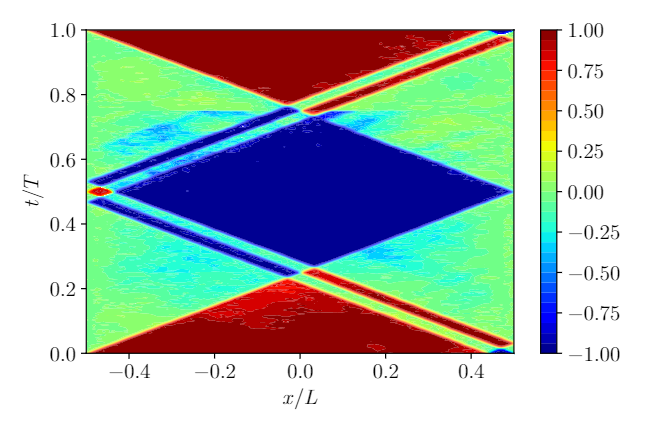}
        \caption{CRAN prediction}
    \end{subfigure}
    \hfill
    \begin{subfigure}[b]{.48\textwidth}
    \centering
    \includegraphics[width=\textwidth]{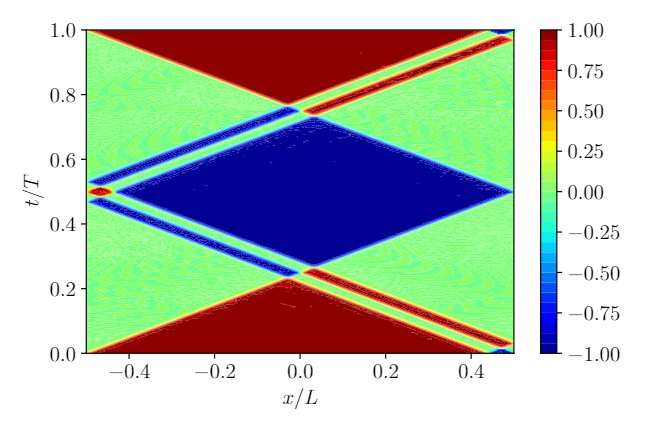}
         \caption{True solution}
    \end{subfigure}
\caption{Comparison of 1D wave propagation predictions from CRAN with true solutions for: $x_s/L = 0.4375$}
\label{fig:comp_1D_discont_xsp4375}
\end{figure}

The predicted and target solutions were also compared for another case, $x_s=0.4375L$. The results are shown in a $x-t$ plane in Fig. \ref{fig:comp_1D_discont_xsp4375}. Similar to the solutions for $x_s=-0.0625L$, we see some differences in the predicted magnitudes over the whole domain at some time steps over the period. However, we again observe that the wave propagation characteristics are accurately represented. The SSIM is a function of not only the structural similarity of the images but also of the luminance and contrast. For the present solution, luminance and contrast can be associated with differences in the magnitude of the target and predicted images, whereas the shape of the characteristics can be associated with the structural quality. For the present case, SSIM is 0.84. Based on the comparison of the solutions in Fig. \ref{fig:comp_1D_discont_xsp4375}, one can state that such reduction from 1 is mostly due to the differences in the magnitude than the structural quality of the images. Since the characteristics are accurately captured by the CRAN for both cases, we may safely conclude the CRAN was able to demonstrate generalized learning of the wave propagation physics for spatially distributed discontinuities. Overall, for all the seven $x_s$ cases not observed by the network during training (Fig. \ref{fig:SSIM_val_test}), CRAN could predict a mean SSIM of 0.90.

\section{\label{sec:7} Conclusions}
In this article, we provide an assessment of the convolutional autoencoder recurrent network (CRAN) as a data-driven model for scalable and generalized learning of wave propagation phenomenon. The CRAN is a composite deep learning model consisting of convolutional autoencoder (CAE) for learning low-dimensional system representation and a long short-term memory (LSTM) recurrent neural network for the system evolution in low dimension.  

We showed that projection-based approaches like POD suffer from slowly decaying Kolmogorov $n$-width when reducing high-dimensional wave propagation data arising from discontinuous initial conditions. However, the CAE could efficiently reduce the high-dimensional physical data with an almost three times higher accuracy than POD. The superior efficiency of the CAE is explained by its ability to learn basis functions directly resembling discontinuous wave characteristics whereas the POD modes lie in Fourier subspace. Also, the scale separation priors of CAE's pooling operations enable it to selectively eliminate low-relevance resolutions from the multi-scale wave dynamics, which further enhances its efficiency to reduce high-dimensional wave propagation data with discontinuities.

To enable autoregressive prediction of wave propagation in low-dimensions via the composite CRAN model, the encoder and decoder of the CAE were connected to an LSTM recurrent neural network. In this regard, the system evolution prediction of popularly used plain LSTM networks was compared with the AR-LSTM and SS-LSTM networks, which have autoregressive modeling capability. The results indicated the poor performance of the plain LSTM, while the SS-LSTM showed the best autoregressive prediction performance. 

The numerical experiments performed here demonstrated the generalized prediction capability of CRAN for wave propagation with spatially distributed discontinuous initial conditions over desired time horizons, while operating on low dimensions. Such generalizarion was indicated by the 90\% mean SSIM shown by the CRAN predictions for out-of-training cases when compared to the true solutions. Furthermore, the pointwise $L_1$ errors of the CRAN predictions usually lied within 10\% for most of the out-of-training cases. Most importantly, the predicted wave propagation closely followed the wave characteristics of the target solutions. Such generalized learning of the CRAN can be attributed to its convolutional encoder and decoders, which demonstrated generalized learning and classification of wave characteristics with spatially varying discontinuous initial conditions. This was indicated by the accurate representation of multi-resolution wave characteristics of out-of-training cases in the feature maps of various layers of the CRAN's trained convolutional encoder. Characteristics represent wave kinematics and their generalized classification and prediction proves the physical consistency of the CRAN's learning procedure.

Here we not only present numerical experiments to demonstrate the generalized and scalable learning wave propagation via the CRAN, but we also physically interpret why the CRAN network is able to generalize and how its convolutional autoencoder achieves dimension reduction. Such interpretation is achieved by connecting the geometric priors of convolutional neural networks to the feature maps of the convolutional autoencoder.

%\begin{figure}[ht]
%\centering
%    \begin{subfigure}[b]{.48\textwidth}
%        \centering
%        \includegraphics[width=\textwidth]{figures/CAN10infmapmp1ls6.pdf}
%        \caption{$f=6$}
%    \end{subfigure}
%    \hfill
%    \begin{subfigure}[b]{.48\textwidth}
%    \centering
%    \includegraphics[width=\textwidth]{figures/CAN10infmapmp1ls16.pdf}
%         \caption{$f=16$}
%    \end{subfigure}
%\caption{...}
%\label{fig:CAN10infmapemp1ls6_16}
%\end{figure}

\appendix
\section{Network parameters} 
\begin{table}[ht]
\caption{Convolutional encoder parameters}
\centering
\begin{center}
\begin{tabular}{lccccc} \hline
Layer \# & Layer type & 
    \begin{tabular}{@{}c@{}}Output \\ dimension\end{tabular}
& Kernel size & 
    \begin{tabular}{@{}c@{}}\# filters/ \\ \# neurons\end{tabular} 
& Stride \\
\hline
 & Input & $\left(16,1025,1\right)$ & - & - & - \\
1 & Convolution 1D & $\left(16,1020,16\right)$ & $\left(6\right)$ & 16 & 1 \\
  & Max Pool 1D & $\left(16,510,16\right)$ & $\left(2\right)$ & - & - \\
2 & Convolution 1D & $\left(16,506,32\right)$ & $\left(5\right)$ & 32 & 1 \\
 & Max Pool 1D & $\left(16,253,32\right)$ & $\left(2\right)$ & - & - \\
3 & Convolution 1D & $\left(16,250,48\right)$ & $\left(4\right)$ & 48 & 1 \\
 & Flatten & $\left(16,12000\right)$ & - & - & - \\
4 & Fully Connected & $\left(16,64\right)$ & - & 64 & - \\
\hline
\end{tabular}
\end{center}
\label{tab:CRAN_enc_1dwave}
\end{table}

\begin{table}[ht]
\caption{LSTM parameters}
\centering
\begin{center}
\begin{tabular}{lccc}\hline
Layer \# & Layer type & 
    \begin{tabular}{@{}c@{}}Output \\ dimension\end{tabular}
& \# neurons \\
\hline
 & Input & $\left(16,64\right)$ & - \\
5 & LSTM & $\left(16,256\right)$ & 256 \\
 & Fully Connected & $\left(1024\right)$ & 1024 \\
 & Reshape & $\left(16,64\right)$ & - \\
\hline
\end{tabular}
\end{center}
\label{tab:CRAN_prop_1dwave}
\end{table}

\begin{table}[ht]
\caption{Convolutional decoder parameters}
\centering
\begin{center}
\begin{tabular}{lccccc}
Layer \# & Layer type & 
    \begin{tabular}{@{}c@{}}Output \\ dimension\end{tabular}
& Kernel size & 
    \begin{tabular}{@{}c@{}}\# filters/ \\ \# neurons\end{tabular} 
& Stride \\
\hline
 & Input & $\left(16,64,1\right)$ & - & - & - \\
6 & Fully Connected & $\left(16,12000\right)$ & - & 12000 & - \\
 & Reshape & $\left(16,250,48\right)$ & - & - & - \\
7 & Convolution 1D Transpose & $\left(16,253,64\right)$ & $\left(4\right)$ & 48 & 1 \\
 & Upsampling 1D & $\left(16,506,48\right)$ & $\left(2\right)$ & - & - \\
8 & Convolution 1D Transpose & $\left(16,510,16\right)$ & $\left(5\right)$ & 32 & 1 \\
 & Upsampling 1D & $\left(16,1020,16\right)$ & $\left(2\right)$ & - & - \\
9 & Convolution 1D Transpose & $\left(16,1025,1\right)$ & $\left(6\right)$ & 16 & 1 \\
\hline
\end{tabular}
\end{center}
\label{tab:CRAN_dec_1dwave}
\end{table}

\section*{Acknowledgments}
This research was supported by the Natural Sciences and Engineering Research Council of Canada (NSERC) [grant number IRCPJ 550069-19]. We would also like to acknowledge that the GPU facilities at the Digital Research Alliance of Canada clusters were used for the training of our deep learning models.

\bibliographystyle{siamplain}
\bibliography{references}

\end{document}